\documentclass{aa}  
\usepackage[utf8]{inputenc}

\usepackage{natbib,twoopt}
\usepackage[colorlinks =true, urlcolor =blue, linkcolor =blue, citecolor =blue]{hyperref} 
\bibpunct{(}{)}{;}{a}{}{,}             
\makeatletter
  \newcommandtwoopt{\citeads}[3][][]{\href{http://adsabs.harvard.edu/abs/#3}%
    {\def\hyper@linkstart##1##2{}%
     \let\hyper@linkend\@empty\citealp[#1][#2]{#3}}}
  \newcommandtwoopt{\citepads}[3][][]{\href{http://adsabs.harvard.edu/abs/#3}%
    {\def\hyper@linkstart##1##2{}%
     \let\hyper@linkend\@empty\citep[#1][#2]{#3}}}
  \newcommandtwoopt{\citetads}[3][][]{\href{http://adsabs.harvard.edu/abs/#3}%
    {\def\hyper@linkstart##1##2{}%
     \let\hyper@linkend\@empty\citet[#1][#2]{#3}}}
  \newcommandtwoopt{\citeyearads}[3][][]%
    {\href{http://adsabs.harvard.edu/abs/#3}
    {\def\hyper@linkstart##1##2{}%
     \let\hyper@linkend\@empty\citeyear[#1][#2]{#3}}}
\makeatother

\usepackage{longtable}
\usepackage{graphicx}
\usepackage{txfonts}

\usepackage{tikz,xcolor,hyperref}

\definecolor{lime}{HTML}{A6CE39}
\DeclareRobustCommand{\orcidicon}{
	\begin{tikzpicture}
	\draw[lime, fill =lime] (0,0) 
	circle [radius =0.16] 
	node[white] {{\fontfamily{qag}\selectfont \tiny ID}};
	\draw[white, fill =white] (-0.0625,0.095) 
	circle [radius =0.007];
	\end{tikzpicture}
	\hspace{-2mm}
}

\foreach \x in {A, ..., Z}{\expandafter\xdef\csname orcid\x\endcsname{\noexpand\href{https://orcid.org/\csname orcidauthor\x\endcsname}
			{\noexpand\orcidicon}}
}

\begin{document}

   \title{Integral field spectroscopy of the planetary nebula NGC\,3242 and the puzzling nature of its low ionization structures }

   \subtitle{}

   \author{L. Konstantinou\inst{1,2}{\orcidD{}}, S. Akras\inst{1}{\orcidA{}},  J. Garcia-Rojas\inst{3,4} {\orcidE{}}, K. Bouvis\inst{1,2}{\orcidC{}}, D.R. Gon\c{c}alves\inst{5}{\orcidJ{}}, H. Monteiro\inst{6,7}{\orcidB{}}, P. Boumis\inst{1}{\orcidG{}}, M. B. Mari\inst{8}{\orcidF{}}, I. Aleman \inst{9}{\orcidI{}} \and 
          A. Monreal-Ibero\inst{10}{\orcidH{}}}
    \authorrunning{L. Konstantinou et al.}
   \institute{Institute for Astronomy, Astrophysics, Space Applications and Remote Sensing, National Observatory of Athens,
              GR 15236 Penteli, Greece
              \email{lydiakonst@gmail.com} 
         \and
            Department of Physics, University of Patras, Patras, 26504 Rio, Greece    
         \and 
             Instituto de Astrof\'isica de Canarias, E-38205 La Laguna, Tenerife, Spain
         \and
             Departamento de Astrof\'isica, Universidad de La Laguna, E-38206 La Laguna, Tenerife, Spain
             \and
             Observat\'orio do Valongo, Universidade Federal do Rio de Janeiro, Ladeira Pedro Antonio 43, 20080-090, Rio de Janeiro, Brazil
              \and 
            School of Physics and Astronomy, Cardiff University, Queen's Buildings, The Parade, Cardiff CF24 3AA, UK
             \and
             Instituto de F\'{i}sica e Qu\'{i}mica, Universidade Federal de Itajuba, Av. BPS 1303-Pinheirinho, 37500-903, \'{I}tajuba, Brazil 
             \and
             Observatorio Astronómico de Córdoba. Universidad Nacional de Córdoba, Laprida 854, Córdoba, Argentina
             \and 
            Laborat\'{o}rio Nacional de Astrof\'{i}sica, Rua dos Estados Unidos, 154, Bairro das Na\c{c}\~{o}es, Itajub\'{a}, MG, 37504-365, Brazil
             \and
             Leiden Observatory, Leiden University, P.O. Box 9513, 2300 RA Leiden, The Netherlands
}

   \date{Received XXXXX, XXXX; accepted XXXXXX, XXXX}

 
  \abstract
   {The physico-chemical properties of the planetary nebula (PN) NGC\,3242 are investigated in both 1D and 2D, using Integral Field Unit (IFU) data. This PN has a complex morphology with multiple shells and contains a pair of structures with a lower degree of ionization compared to the main nebular components. These structures are known as low ionization structures (LISs), and their origin is still a mystery.}
    { With the capabilities provided by IFU spectroscopy, we aim to gain a better understanding of the behavior of nebular properties in the LISs, and examine the spatial distribution of physico-chemical parameters in NGC\,3242.}
   {Data from the Multi Unit Spectroscopic Explorer (MUSE) at the Very Large Telescope (VLT) were used in order to perform a spatially resolved physico-chemical analysis of NGC\,3242 both in 2D, through the analysis of emission line maps, as well as in 1D, simulating long-slit spectroscopy, with pseudo-slits.}
   {Through the deeper investigation of MUSE data, we detect new structures perpendicular to the pair of LISs of NGC\,3242, which are mainly seen in the light of {[S~\sc iii]} and {[N~\sc ii]}. In addition, two arc-like structures are revealed. Moreover, an inner jet-like structure is found through its [Fe~{\sc iii}] emission. The interaction of the jet with the rim may be related to the formation of knots and blobs. The higher value of $T_{\rm e}$, is estimated from the {[S~\sc iii]} diagnostic lines, followed by $T_{\rm e}$ ({[N~\sc ii]}), $T_{\rm e}$({H~\sc i}) and finally $T_{\rm e}$ ({He~\sc i}). In all cases, $T_{\rm e}$ is higher at the inner nebular structures. Regarding electron density, $n\rm _e$, is lower at the LISs, while an increase is observed at the nebular rim. Diagnostic diagrams confirm that NGC\,3242 is a highly ionized nebula.   
   Moreover, the MUSE data unveiled for the first time in this PN, the atomic line [C~{\sc i}] $\lambda$8727, primarily emitted from the LISs. This finding suggests that these structures may consist of a molecular core surrounded by neutral and ionized gas.}
   {}

    \titlerunning{Integral field spectroscopy of NGC\,3242 and the puzzling nature of its low ionization structures}
    \maketitle

\section{Introduction}

As low- to intermediate-mass stars reach the end of their lives, strong stellar winds occur which lead to the formation of an ionized gas region, known as Planetary Nebula (PN). The complicated morphology of PNe observed in most cases (e.g., irre\-gu\-lar, point symmetric, asymmetric, and bipolar) cannot be explained from the Generalized Interacting Stellar Wind (ISW) model \citep{Balick1987, Icke_1988}. So, many studies have proposed different scenarios regarding their formation mechanisms \citep[see][for a review]{Balick_review}. 
The main morphological structures of PNe are the bright rim, the shells, and the halo. In addition, "rings" have been identified embedded in the halos of some PNe caused by different episodes of mass loss during the Asymptotic Giant Branch (AGB) phase \citep{Coradi2003}.

Another open question in the study of PNe is the formation of small-scale structures with a lower ionization degree compared to the surrounding gas. These structures are known as low ionization structures (LISs), as the lines emitted from them are mainly low ionization ones such as [N~{\sc ii}], [S~{\sc ii}], and [O~{\sc i}] \citep{Corradi_1996, Balick_LISs, 2001_Denise_LIS, LIS_Akras_2015, Belen_III}. LISs are classified according to their morphology as knots, filaments, jets, and jet-like systems \citep{2001_Denise_LIS}. Additionally, they are further divided into slow-moving low-ionization emitting regions or SLOWERS \citep{SLOWERS} and fast low-ionization emission regions or FLIERS \citep{FLIERS}. The first
category consists of LISs with radial velocities similar to the
other nebular structures (e.g., NGC\,7662, K\,1-2, and Wray\,17-1), while the second one includes those with velocities of $\pm$25~$km\,s^{-1}$ \citep[or even higher in some cases][]{FLIERS} with respect to their surrounding gas (e.g., NGC\,3242, NGC\,7009, and NGC\,6543). Both the nature and the mechanisms that led to the formation of LISs, are not fully understood. Several scenarios have been proposed to explain their origin, but none of them seem to cover all types of LIS.

Here, we investigate the spatial distribution of the physico-chemical properties of NGC\,3242, with a main focus on its LISs. To this end, MUSE data are employed in conjunction with the {\sc sa\-te\-lli\-te} code, developed by \cite{SATELLITE,SATELLITE_case_studies}. NGC\,3242, also known as Ghost of Jupiter, is located appro\-xi\-ma\-te\-ly 1\,279$^{+63}_{-62}$ pc away, as it was estimated from GAIA mission \citep{Distance}. Its central star (CS) has an effective temperature of $\sim$80\,000 K and a radius of 0.14\,R$_\odot$ \citep{Pottasch}. As for its physico-chemical properties, the electron temperature ($T_{\rm e}$) is around 11\,000 K calculated either from {[O~\sc iii]} or [N~{\sc ii}] emission line ratios, while the electron density ($n_{\rm e}$) is about 10$^3$ cm$^{-3}$ based on [S~{\sc ii}] and [Cl~{\sc iii}] emission line ratios \citep{Balick1993,Monreal_Ibero_2005_ngc3242, Krabe_Copetti_2005, Pottasch, Monteiro2013, Miller_2016, Pottasch}.

\begin{figure*}
    \centering
    \includegraphics[width =0.8\textwidth]{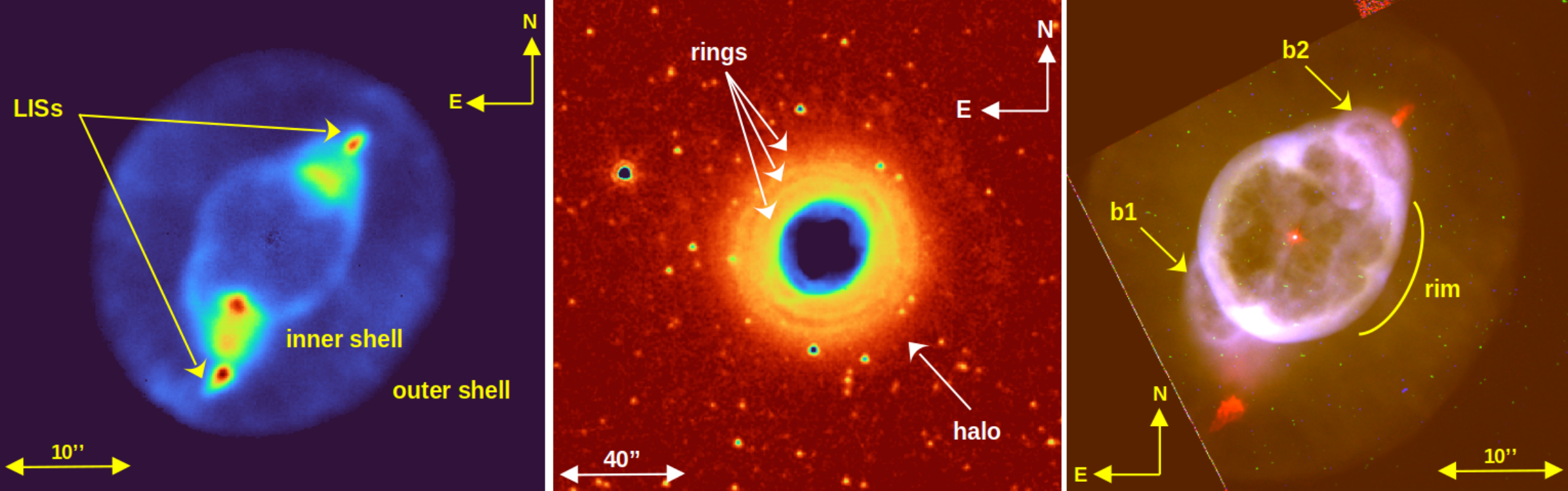}
    \caption{Structures of NGC\,3242. The left panel shows the {[N~\sc ii]} $\lambda$6584 map from MUSE datacubes where the LISs and the nebular inner and outer shells are indicated. The middle panel illustrates the ratio 8.0/4.5 of IRAC bands, where three rings and the halo are visible. The final figure is an RGB image (R: {[N~\sc ii]} $\lambda$6584, G: H$\rm\alpha$, B: {[O~\sc iii]} $\lambda$5007) from HST data, which shows the rim and the blobs.}
    \label{PN_struct}
\end{figure*}

In terms of morphology, NGC\,3242 is a multiple shell PN that contains a pair of LISs. Its inner shell is elliptical with size 28$^{\prime\prime}$ $\times$ 20$^{\prime\prime}$ and contains the rim, while the outer shell is 46$^{\prime\prime}$~$\times$~40$^{\prime\prime}$ and almost elliptical \citep{Monteiro2013}. In addition to  the inner large scale structures, microstructures are also identified in the form of blobs and LISs. More specifically, two blobs \citep[b1 and b2 in fig.1 from][]{Morphokinematcs} are resolved NW and SE from the CS, respectively, while the LISs, also identified as FLIERS, appear between the two shells \citep{Meaburn_2000, Morphokinematcs}, along the NW-SE orientation \citep{Coradi2003}. The velocities of the SE and the NW LISs are 22 and $-28$ km\,s$^{-1}$ (relative to the systemic velocity, Vsys = (-6.6$\pm$1) km\,s$^{-1}$), respectively, estimated from high resolution spectra obtained with the Manchester Echelle Spectrograph \citep[MES, see][]{Morphokinematcs}. The external shell is further surrounded by an extended halo. The latter is identified as a circular or slightly elliptical AGB halo, representing the final product of the thermal pulses at the end of the AGB phase \citep{Coradi2003}. At least three ring-like structures are observed within this halo via Hubble Space Telescope (HST) and Spitzer Space Telescope data \citep{Morphokinematcs, Monreal_Ibero_2005_ngc3242, Philips}. In addition, another extended ionized halo has been identified surrounding the first one \citep{Coradi2003}. Mid-infrared image data from the InfraRed Array Camera (IRAC) on Spitzer revealed that the external halo is also clumpy, possibly due to its fragmentation through Rayleigh-Taylor instabilities \citep{Ramos_Phillips_2009}. The aforementioned structures of NGC\,3242 are illustrated in Fig.~\ref{PN_struct}. More precisely, the left-panel of Fig.~\ref{PN_struct} shows the emission map of {[N~\sc ii]} $\lambda$6584 (see Sect.~\ref{sec:methodology}, for details), labelling the nebular shells and the pair of LISs of NGC\,3242. The middle and right panels were created using archival data from the Spitzer Space Telescope (programme: 1427, PI: calibration IRS)  and the HST (programme: 6117, PI: Balick). The middle panel represents the ratio 8.0/4.5 of the Spitzer's Infrared Array Camera (IRAC) bands. In the same image, the outer nebular structures are indicated (rings and halo) in the wider field of view of the instrument (5.2$^\prime$$\times$5.2$^\prime$). Finally, an RGB image was created resulting from the combination of {[N~\sc ii]}: F658N (Red),  H$\rm\alpha$: F656N (Green), {[O~\sc iii]}: F502N (Blue) filters available on Wide Field Planetary Camera 2 (WFPC2) of HST. In this final panel, the rim and the blobs are labelled.

In general, several studies have investigated the variation in physical and chemical properties throughout PNe with a particular focus in their LISs, e.g., \cite{Hajian, Balick_LISs, 2001_Denise_LIS, Goncalves_2009, LIS_Akras_2015, Belen_III, Belen_II}. According to these studies, there is no significant variation in $T_{\rm e}$ between the main ne\-bu\-la and the LISs. On the contrary, $n_{\rm e}$ seems to decrease at the position of the LISs. This is the opposite result of what is expected from the theoretical scenarios on their formation, which predict that LISs should be denser than their surrounding enviroment \citep{model_2001, model_2008, model_2020}. A possible explanation for this disagreement is that LISs contain a molecular gas component, which is not considered when $n_{\rm e}$ is computed. Indeed, observations in several PNe have revealed the existence of molecular hydrogen in LISs \citep[e.g.,][]{Fang_2015, Fang_2018, AKras_2017, LIS_H2_Akras}, and in the recently discovered clumps of NGC\,3587 by \cite{ngc3587}.

In Sect.~\ref{sec:methodology} we describe our approach and methodology. The results from the 1D and 2D analysis are presented in Sect.~\ref{sec:results}. In Sect.~\ref{sec:discussion} we discuss our results, and in Sect.~\ref{sec:summary} we summarize the main conclusions of this study.

\section{Methodology}
\label{sec:methodology}

Integral Field Spectroscopy (IFS) offers the unique opportunity of simultaneously producing spatially resolved spectrum, in two dimensions. In terms of PNe studies, this approach offers valuable insights into the spatial distribution of the emitted nebular spectra. NGC\,3242 was first studied with the Integral field unit (IFU) Visible Multi-Object Spectrograph (VIMOS) by \cite{Monreal_Ibero_2005_ngc3242} and some properties of its halo were identified. Later on, \cite{Monteiro2013} used data from the same instrument for a more extended investigation of the physico-chemical properties of this PN. In this work, archival data from MUSE IFU are used for the spectroscopic characterization of NGC\,3242 (programme: 097.D-0241(A), PI: Corradi R.L.M.). MUSE has a wider field of view and better spectral and spatial resolution than VIMOS. In its nominal Wide-Field Mode (WFM-NOAO-N), MUSE covers a wavelength range from 480 to 930 nm, with a field of view 1$^\prime$ $\times$ 1$^\prime$, and a pixel scale of 0.2$^{\prime\prime}$ per pixel. Details on the reduction steps of the data can be found in \citet{ADF_recomb}. Table~\ref{Tab:Obs} summarizes the log of observations. For our analysis, two data cubes were used, one with an exposure time of 10~s and a deeper one with a total exposure time of 900~s. From these datacubes, we extracted emission line maps of some optical emission lines, typically observed in PNe and the error maps as produced by the fitting. All the fluxes and the associated errors were measured as described in Sect. 3 of \citet{ADF_recomb}. The brightest emission lines (H$\rm\alpha$, H$\rm\beta$ and [O {\sc iii}] $\lambda$$\lambda$4959, 5007) maps were extracted from the 10\,s datacube, as they are saturated in the longer exposure datacube. Overall, 23 emission line maps were extracted from MUSE datacubes. In Table \ref{line_fluxes} the observed and de-reddened line fluxes are listed, integrated from the whole PN (see the pseudo-slit dimension on the same table).

\begin{table}
\caption{Observing log of the MUSE observations.}
\label{Tab:Obs}
\begin{tabular}{|ccccc|}
\hline
UT Start & n & Exp &  Airm. & Seeing \\
& & (s) &  & ($"$) \\\hline
Target: NGC\,3242 & \multicolumn{4}{c|}{Mode:  WFM--NOAO-N}\\ \hline
 2016-07-06\hspace{1.5mm}23:31:26.133&1/9&10.0&1.50&2.0\\
 2016-07-06\hspace{1.5mm}23:33:35.638&2/9&60.0&1.50&1.5\\
 2016-07-06\hspace{1.5mm}23:36:52.233&3/9&180.0&1.53&1.5\\
 2016-07-06\hspace{1.5mm}23:41:23.196&4/9$^{\rm a}$&60.0&1.55&1.0\\
 2016-07-06\hspace{1.5mm}23:44:32.674&5/9&180.0&1.58&0.9\\
 2016-07-06\hspace{1.5mm}23:49:29.663&6/9&180.0&1.62&1.1\\
 2016-07-06\hspace{1.5mm}23:54:00.244&7/9$^{\rm a}$&60.0&1.65&1.2\\
 2016-07-06\hspace{1.5mm}23:57:09.890&8/9&180.0&1.69&1.2\\
 2016-07-07\hspace{1.5mm}00:02:05.466&9/9&180.0&1.73&1.0\\
\hline
\end{tabular}
\tablefoot{
\tablefoottext{a} {Sky frames were taken 5 arcmin away from the object to ensure that there was no nebular contamination.}
}
\end{table}

\begin{figure*}[h!]
    \centering{
    \null\hfill
    \caption*{}
    \hfill
    \subfloat{\includegraphics[width =0.47\textwidth]{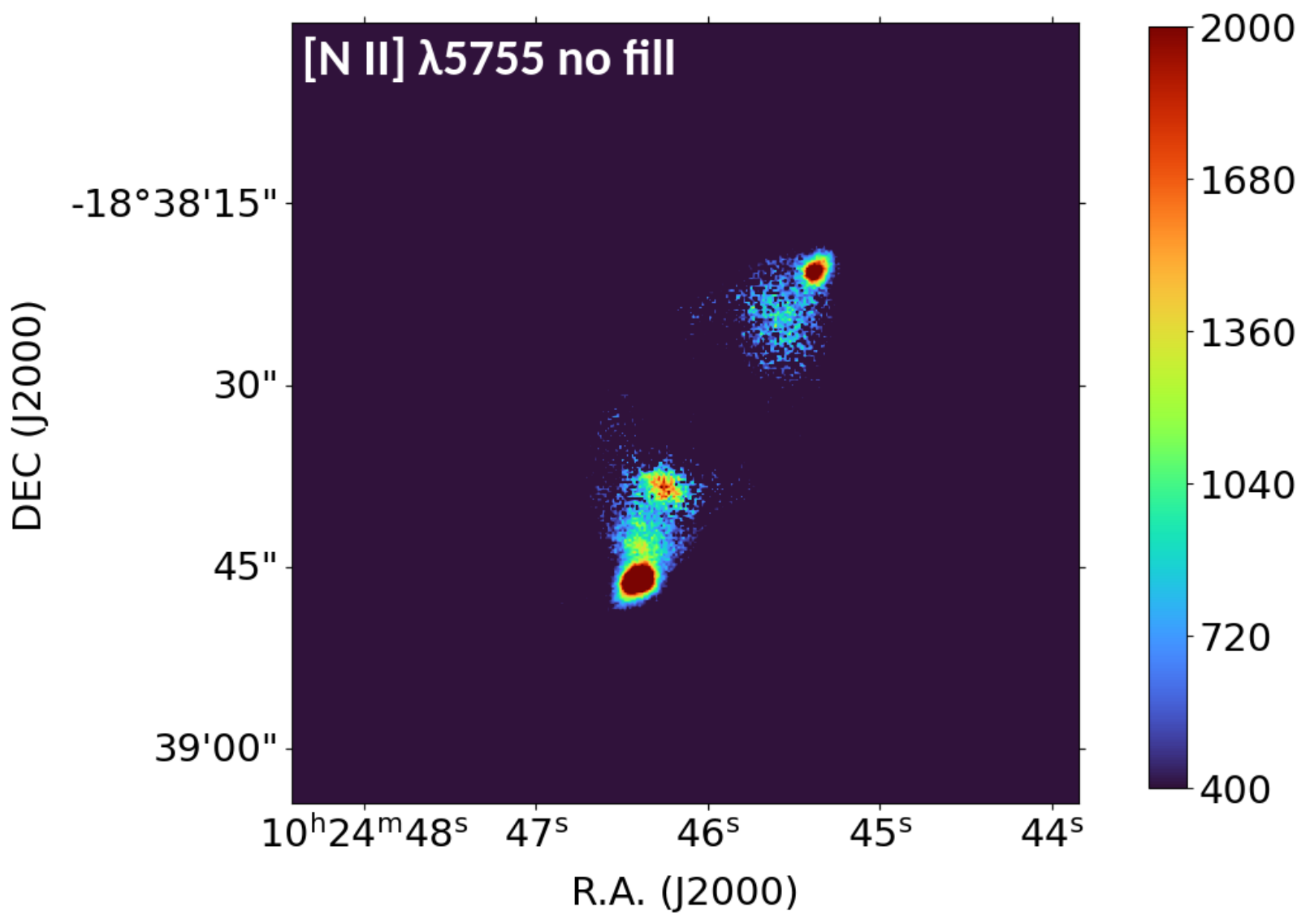}}
    \hfill
     \subfloat{\includegraphics[width =0.47\textwidth]{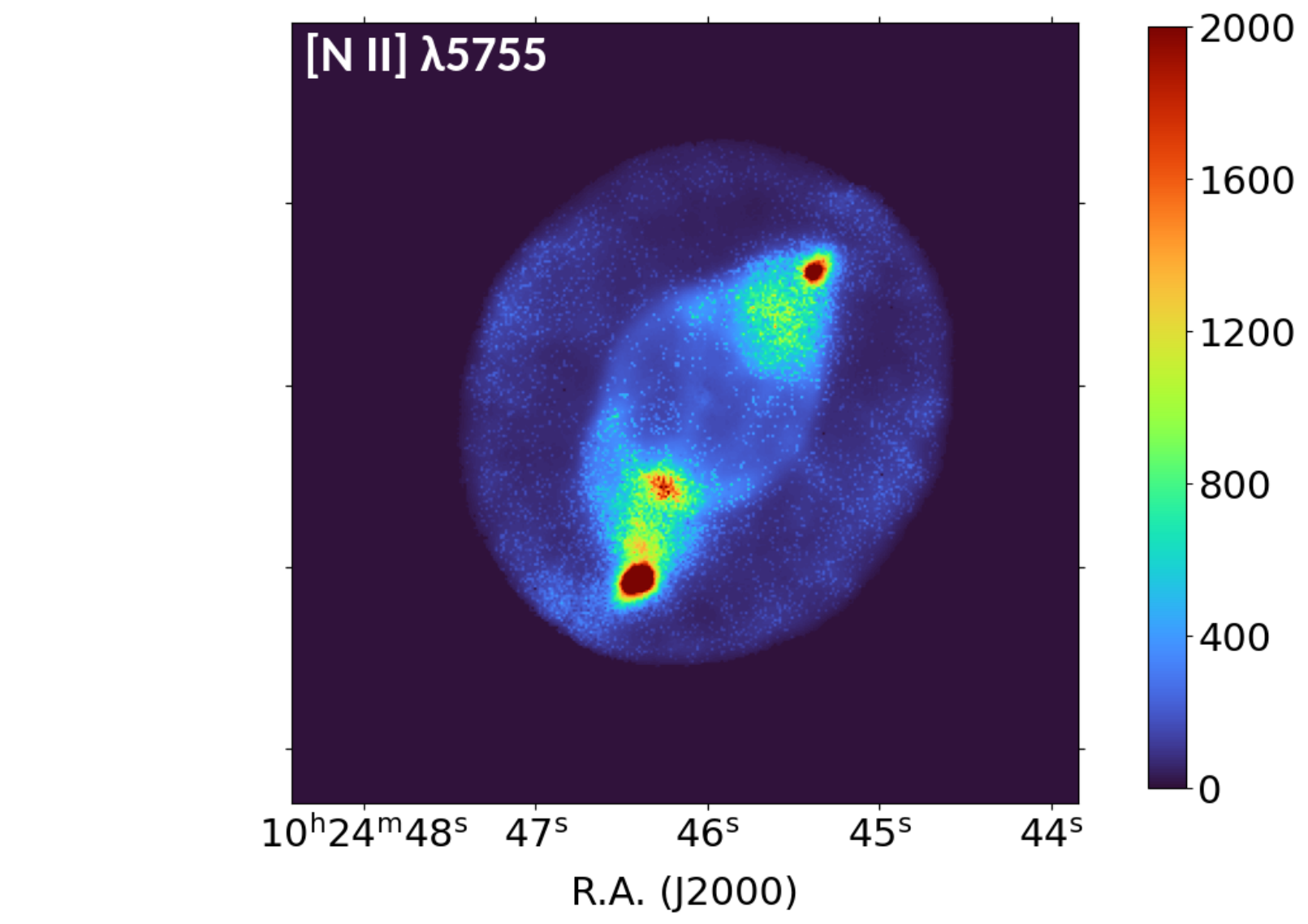}}
    \null\hfill
    }
\caption{Emission line maps of [N~{\sc ii}] $\lambda$5755. The left panel illustrates the map after masking the pixels with low S/N, while in the right one is the final map that we used. The colorbar represents pixel values (in terms of flux units $\times$10$^{-20}$ erg\,s$^{-1}$\,cm$^{-2}$\,spaxel$^{-1}$)}
\label{N2}
\end{figure*}

\subsection{Masking on the emission maps}
\label{sec:masks}

To exclude pixels with low low signal-to-noise ratio (S/N), a mask was applied considering F(H$\rm\beta$) > 0.002 $\times$ F(H$\rm\beta$)$_{max}$, where F(H$\rm\beta$)$_{max}$ is the maximum intensity of H$\rm\beta$. A similar mask, but with 5\% of F(H$\rm\beta$)$_{max}$, was applied by \cite{ADF_recomb}. The lower limit was chosen because the LISs are relatively faint in H$\rm\beta$ compared to the rest of the nebula.

At some nebular regions, $T_{\rm e}$ ([N~{\sc ii}]) was unrealistic, due to the low S/N in the measurement of the [N~{\sc ii}] $\lambda$5755 line flux. So, in addition to the previous criterion, another mask was applied to the already corrected for recombination contribution [N~{\sc ii}] $\lambda$5755 diagnostic line map (Sect. \ref{recombination contrib}), to exclude pixel values with low S/N. To this end, the pixels where the ratio of the emission line map to the error map fell below 5, were set equal to zero (see Fig.~\ref{N2}, left panel). 
Then, the zero values in the {[N~\sc ii}] $\lambda$5755 emission map were replaced so the temperature diagnostic ratio {[N~\sc ii}] ((6548+6584)/5755)  remains constant, producing a 2D map with constant $T_{\rm e}$ = 11\,400 K (see Fig.~\ref{N2}, right panel). This $T_{\rm e}$ represents the mean value of the electron temperature at the LISs, where the {[N~\sc ii}] $\lambda$5755 emission line has the higher S/N. We consider this a reasonable approximation, as previous 2D studies of NGC\,3242 found no significant variation in $T_{\rm e}$[N~{\sc ii}] across different nebular regions \citep{Monteiro2013}.

\subsection{{\sc satellite} code}
We use the {\sc satellite} code \citep{SATELLITE,SATELLITE_case_studies} to perform the spectroscopic analysis of NGC\,3242 MUSE data. This code uses IFU data to carry out in-depth spectroscopic characterization of ionized nebulae (PNe, {H~\sc ii} regions or star forming galaxies). {\sc satellite} code has four modules: specific slit analysis, rotational analysis, radial analysis, and 2D analysis tasks. {\sc satellite} makes use of {\sc PyNeb} \citep[version 1.1.18,][]{pyneb, pyneb2_2020}, and the atomic data available, to compute the physical and chemical properties of the gas. The atomic data that were employed in this study are listed in Table \ref{atomic_data}. The emission lines are corrected for extinction, using {\sc PyNeb}, by comparing the estimated ratio H$\alpha$/H$\beta$ with the theoretical one  \citep[H$\alpha$/H$\beta$=2.86 in the low density limit for $T_{\rm e}$=10\,000~K,][]{Osterbrock, Storey1995}. The interstellar extinction law (R$_v$) is a free parameter in {\sc satellite}, and here we employed R$_v$ = 3.1 \citep{Seaton1979, Howarth1983, Cardelli1989}.  Then the nebular physical parameters such as $T_{\rm e}$ and $n_{\rm e}$, as well as the chemical parameters, such as ionic abundances, ionization correction factors (ICFs), total elemental abundances and abundance ratios, can be calculated using any of these modules. The uncertainties of the physico-chemical parameters in the 1D analysis, are estimated based on 100 Monte Carlo simulations\footnote{The errors remain almost invariant, even for a much higher number of Monte Carlo simulations. So, 100 iterations were chosen to achieve a balance between accurate uncertainties and computational runtime \citep[see][]{DelgadoInglada_2015, ADF_recomb, GomezLlanos2024}.}. For each line intensity, random values are generated, using a Gaussian distribution centered on the observed line intensity with a standard deviation equal to its uncertainty. The code uses 35 emission lines, including the 23 that are the object of this study, both optical recombination lines (ORLs) and collisionally excited lines (CELs), typically observed in ionized nebulae. {\sc satellite} can do both 1D and 2D analysis, where in the first case, a number of pseudo-slits is used to simulate slit spectroscopy, hence offering the opportunity of comparing results from IFS with previous studies that have been conducted using long-slit observations.

In the case of the specific slit analysis task, a default number of ten pseudo-slits is used for the spectroscopic characterization of the areas defined by them. This module can be used for the study of specific morphological nebular components. Regarding the rotational analysis, a number of pseudo-slits are simulated in different position angles (P.A.) from 0 to 360 degrees, while in the radial analysis module, a pseudo-slit is used to test the variation of nebular parameters with the distance from the CS. {\sc satellite} also performs spectroscopic analysis in two spatial dimensions, and the main results of this analysis are 2D maps of the nebular physical parameters. At the same time, emission line diagnostics are provided to better distinguish the excitation mechanisms occurring in different PNe structures. More detailed information on the modules, and the spectroscopic analysis that {\sc satellite} code performs, is available in \cite{SATELLITE, SATELLITE_case_studies}.

\begin{table}
    \caption{Atomic data for CELs and ORLs}
    \label{atomic_data}
    \resizebox{.5\textwidth}{!}{

    \begin{centering}
        \begin{tabular}{|lcc|}
        \multicolumn{3}{c}{CELs}\\
    \hline
    Ion & Transition probabilities & Collision strengths \\\hline
     O$^0$ &  \cite{Wiese1996} & \cite{Bhatia1995} \\
     O$^{+}$ &  \cite{Zeippen1982} & \cite{Kisielius2009}\\
     &  \cite{Wiese1996}&\\
     O$^{2+}$ &  \cite{Storey2000} & \cite{Storey2014}\\
        &  \cite{ FroeseFischer2004} & \\
     N$^{0}$ &  \cite{Kaufman1986} & \cite{Peguignot1976}  \\
             & \cite{ Wiese1996}&\cite{Dopita1976} \\
     N$^{+}$ &  \cite{FroeseFischer2004} & \cite{Tayal2011}\\
    S$^{+}$ &  \cite{Rynkun2019} & \cite{Tayal2010}\\
    S$^{2+}$ &  \cite{FroeseFischer2006} & \cite{Tayal1999}\\
    Cl$^{2+}$ & \cite{Rynkun2019} & \cite{Butler1989}\\
     Ar$^{2+}$ &  \cite{Burgos2009} & \cite{Burgos2009}\\ \hline

     \multicolumn{3}{c}{ORLs}\\
      \hline
    \multicolumn{3}{|c|}{Effective Recombination coefficients} \\
    \hline
    H$^{+}$ & \multicolumn{2}{c|}{\cite{Storey1995}}  \\
    He$^{+}$ &  \multicolumn{2}{c|}{\cite{Porter_2012,Porter_2013}} \\
        He$^{2+}$ &   \multicolumn{2}{c|}{\cite{Storey1995}}\\ \hline

    \end{tabular}
    \end{centering}
    }
\end{table}

\section{Results}
\label{sec:results}

\subsection{The low ionization structures}

\begin{figure*}[h!]
    \centering{
    \null\hfill
    \hfill
     \subfloat{\includegraphics[width =0.75\textwidth]{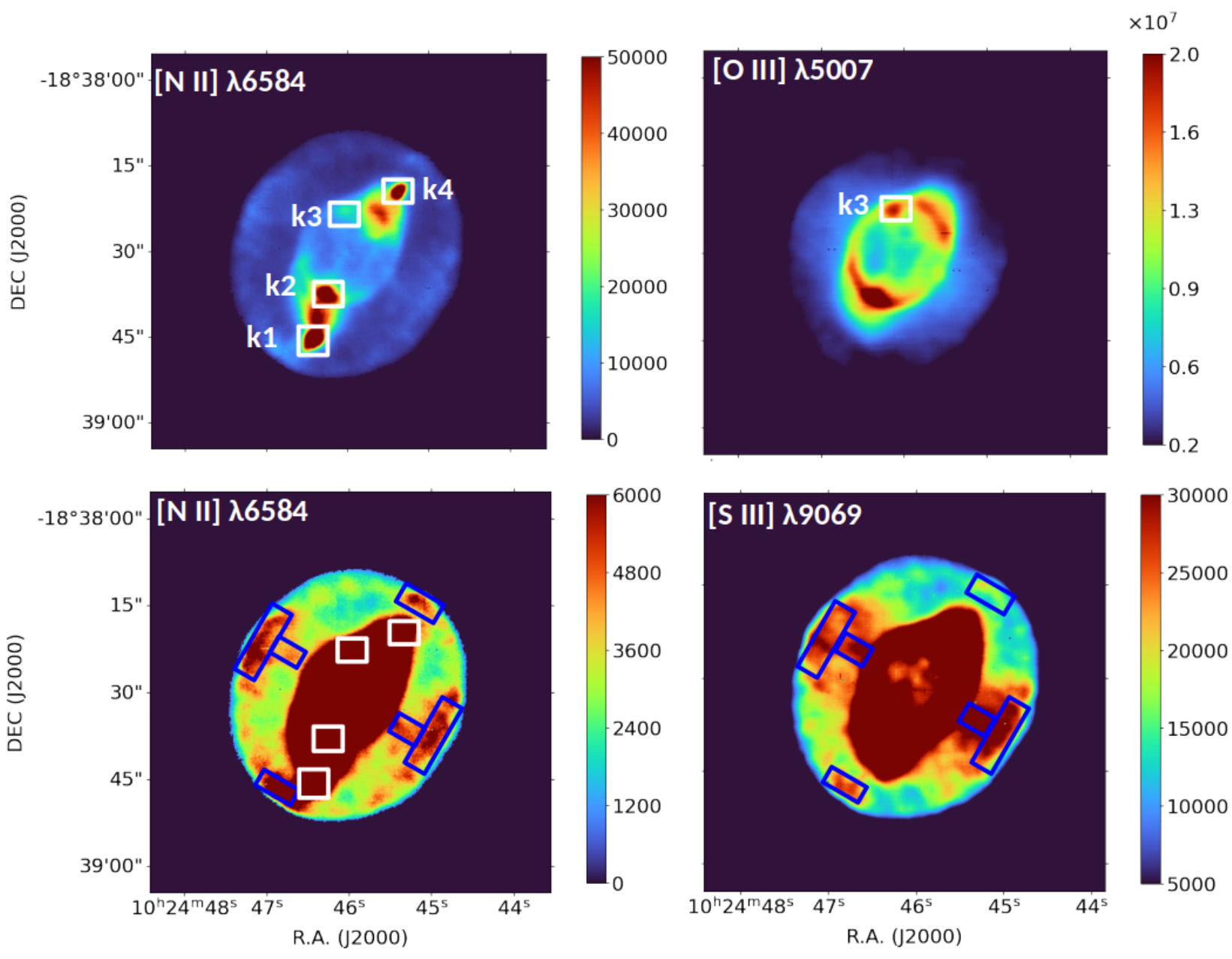}}
    \null\hfill
    }
\caption{Emission line maps of {[N~\sc ii]} $\lambda$6584 (top left and bottom left), {[O~\sc iii]} $\lambda$5007 (top right) and {[S~\sc iii]} $\lambda$9069 (bottom right) for NGC\,3242. 
In the top figures, the four knots are marked. The bottom high-contrast images show the new structures at the nebular shell (blue rectangles), while in the left-bottom figure the four knots are marked as well (white rectangles). The colorbar represent flux values (in units of $\times$10$^{-20}$ erg\,s$^{-1}$\,cm$^{-2}$\,spaxel$^{-1}$).}
\label{emission_maps}
\end{figure*}

In the top-left panel of Fig.~\ref{emission_maps}, we present the {[N~\sc ii]} $\lambda$6584 emission line map. The bright rim and the pair of LISs are prominent in this map. They appear at position angles 160\degr~and 320\degr, and their distances from the CS are 13$^{\prime\prime}$ and 16$^{\prime\prime}$, respectively. A single knot can also be identified near the nebular rim, at a distance of 8$^{\prime\prime}$ from the central star, while {[O~\sc iii]} emission line map (top-right panel) reveals a knot at P.A. = 12\degr~and distance 7$^{\prime\prime}$ from the CS (k2 and k3 in Fig.~\ref{PN_struct}, respectively). \cite{Morphokinematcs} first identified  these knots via imaging with the Harold Johnson Telescope at the San Pedro M\'{a}rtir Observatory (OAN-SPM). In the same study, a fifth knot was found, named as k4b. However, in MUSE data, this knot is not identified. For consistency, knots are marked with the same labels as in the previous work by \cite{Morphokinematcs}.

\begin{figure}[h!]
     \centering
	\begin{subfigure}{2\linewidth}
	    \includegraphics[width =0.52\textwidth]{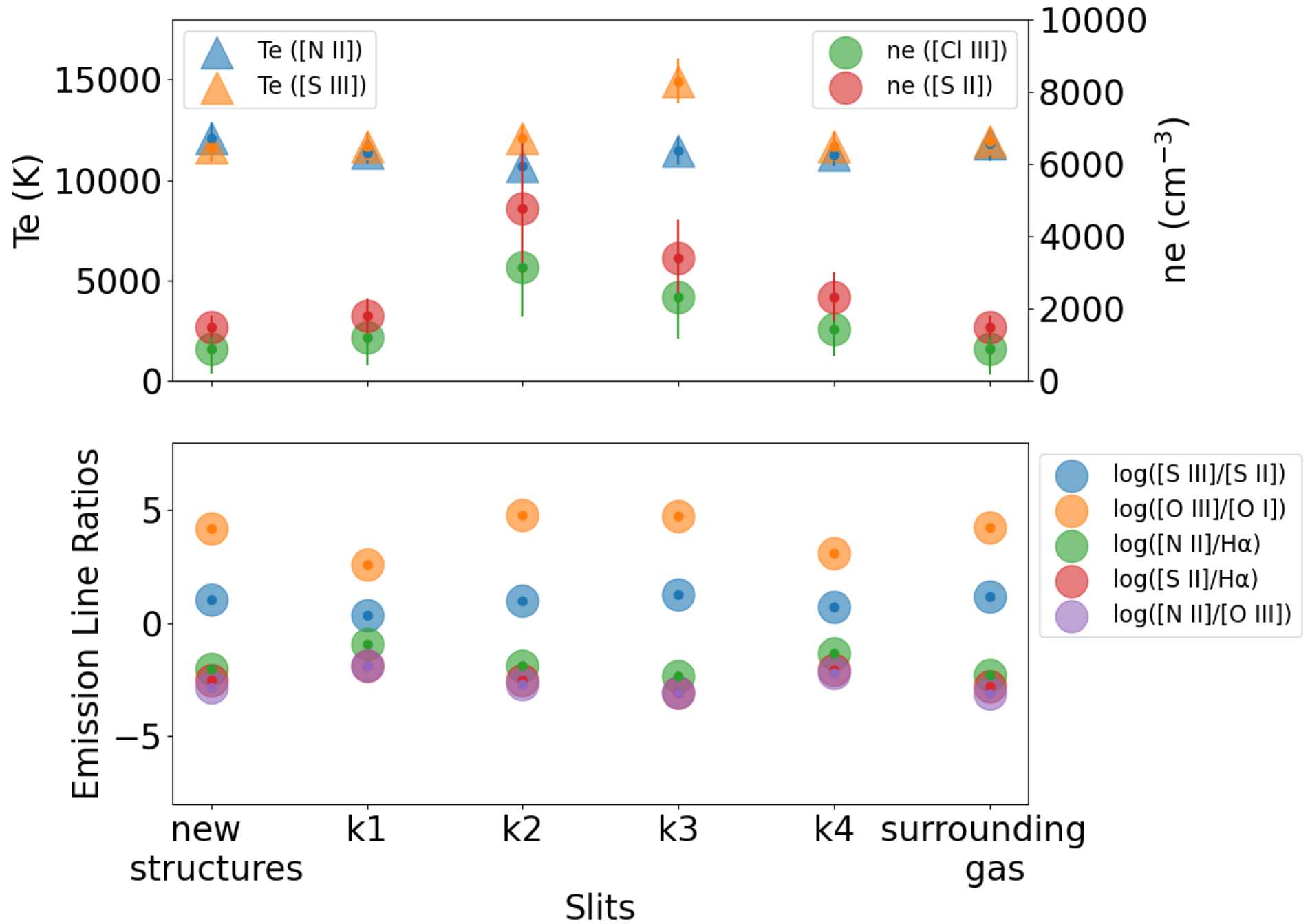}\\
            
	\end{subfigure}
     \caption{The top panel shows the $T_{\rm e}$ and $n_{\rm e}$ for the different structures listed in Table \ref{table_shell_knots}, while the bottom panel shows some emission line ratios for the same structures. In several cases, the error-bars are smaller than the symbols (see Table \ref{table_shell_knots}).}
     \label{ratios_TeNe_knots_sp_slit}
\end{figure}

Using the specific slit analysis module, we investigated the physical properties of the knots. For this analysis, four pseudo-slits were placed at the positions of the southeast (SE) and northwest (NW) LISs, hereafter referred to as k1 and k4, as well as at the positions of knots k2 and k3 (see top panels of Fig.~\ref{emission_maps}). $T_{\rm e}$, $n_{\rm e}$ and some typical emission line ratios are presented in Fig.~\ref{ratios_TeNe_knots_sp_slit} and Table \ref{table_shell_knots}. Emission line ratios, such as log({\sc [S iii]}/{\sc [S ii]}) and log({\sc [O~iii]}/{\sc [O~i]}) \citep{Belen_III}, provide key information on the ionization level of nebular structures. In the case of k1 and k4, the sum of {[S\sc~ii]} emission lines is approximately twice as bright as the sum of {[S\sc~iii]} lines, with log({\sc [S iii]}~(9069+6312)/{\sc [S ii]} (6716+6731)) values of 0.37 and 0.58, respectively. In contrast, k2 and k3 appear to have higher values for the same ratio than the pair of LISs, both around 1. A similar trend is observed for log({\sc [O~iii]}~(4959+5007)/{\sc [O i]} (6300+6363)), with the ratio being about 2 orders of magnitude higher in k2 and k3 compared to k1 and k4. This indicates that knots k2 and k3 are more prominent in high-ionization species, such as S$^{2+}$ and O$^{2+}$, while low-ionization species dominate the emission in k1 and k4. 

Moreover, log({\sc [N ii]} (6548+6584)/H$\alpha$) is almost one order of magnitude higher in k1 and k4 compared to k2 and k3. However, both k2 and k3 appear to be part of the rim, which could influence the estimated flux of low-ionization species. Since we cannot entirely rule out the rim's contribution, it remains possible that k2 and k3 share characteristics with other low-ionization features. Furthermore, $n_{\rm e}$ ([S~{\sc ii]}) is higher in k2 and k3, with mean values $\sim$4\,500 and $\sim$3\,400 cm$^{-3}$, respectively. Since these structures seem to be parts of the nebular rim, we compare our estimations with the rim's $n_{\rm e}$. The density values of the rim close to k2 are $\sim$3\,500~cm$^{-3}$. This suggests that $n_{\rm e}$ in k2 is around 1\,000~cm$^{-3}$ above the rim density. However, it has to be noted that the errors in the estimation of  $n_{\rm e}$ are high. On the contrary, in the case of k3, $n_{\rm e}$ conform with the electron density of the surrounding part of the rim.

In the case of $T_{\rm e}$\,([S~{\sc iii}]), k3 is around 2\,000 K hotter than the other knots, which show a nearly constant $T_{\rm e}$\,([S~{\sc iii}])$\sim$12\,000~K. On the other hand, $T_{\rm e}$\,([N~{\sc ii}]) is $\sim$11\,400~K for k1 and k4, while k2 and k3 exhibit 10\,700~K and 11\,400~K respectively (Fig.~\ref{ratios_TeNe_knots_sp_slit} and Table \ref{table_shell_knots}). However, it should be noted that at the position of k2 and k3, [N~{\sc ii}] $\lambda$5755 emission line is contaminated by recombination, affecting $T_{\rm e}$ ([N~{\sc ii}]) at these regions (see Sect. \ref{recombination contrib}). \cite{Morphokinematcs}, suggested that even if $T_{\rm e}$ is not significantly higher at k2 and k3, they could still have been created from shock material. Here, we propose that they are probably the result of a dynamic interaction at the rim, caused by an inner jet (see Sect. \ref{Iron_emission}).

\begin{figure}[h!]
    \centering{
    \null\hfill
    \caption*{}
    \hfill
    \subfloat{\includegraphics[width =0.45\textwidth]{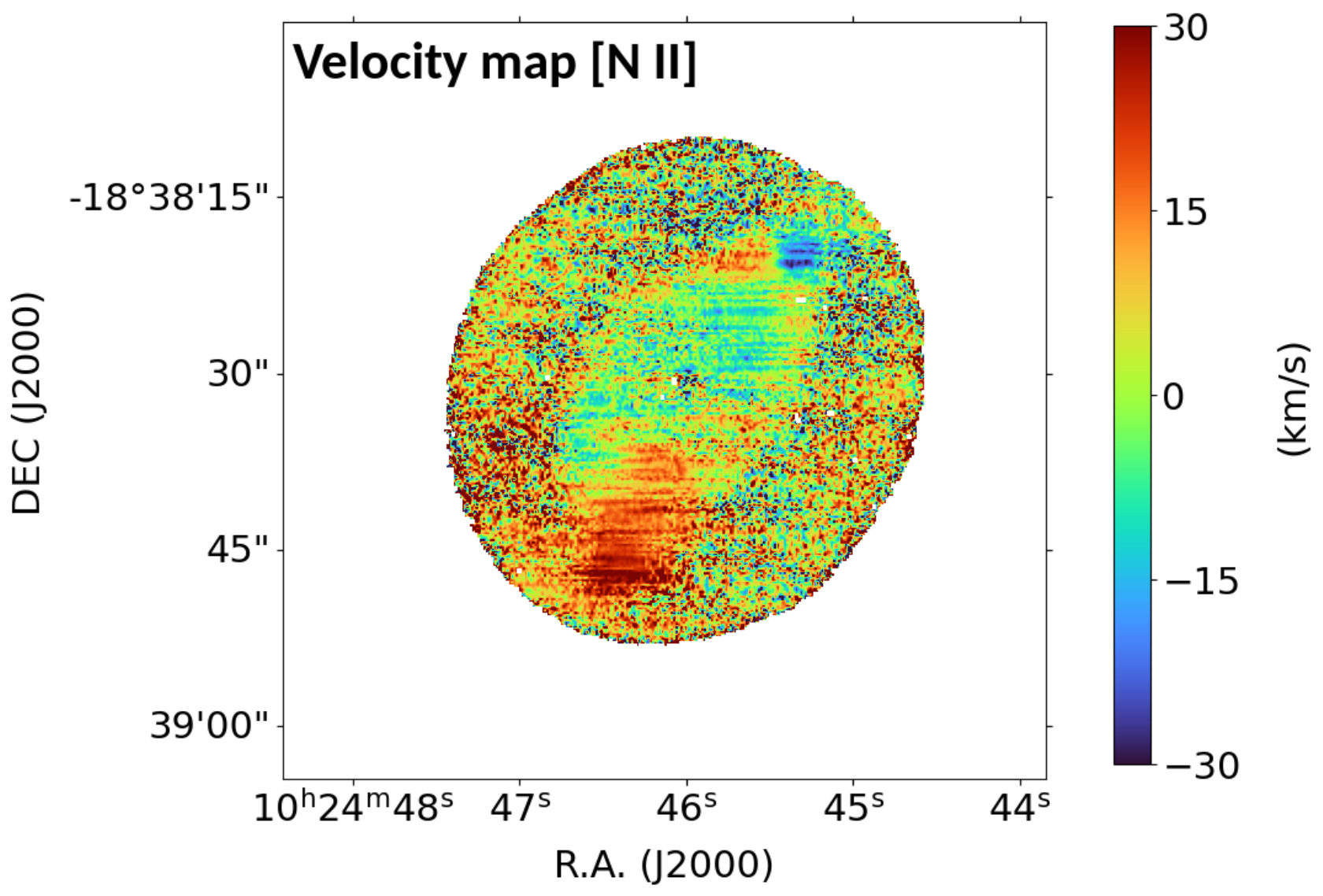}}
    \null\hfill
    }
\caption{Velocity map from {[N~\sc ii]} $\lambda$6584 emission line. The colorbar represent absolute velocities.}
\label{velocity}
\end{figure}

Additionally, 2D velocity maps were constructed based on {[N~\sc ii]} emission lines to examine the kinematics of the LISs. The resolution of MUSE is not sufficient for a thorough kinematic analysis, and the errors of the estimated velocities are high ($\sim$~60~km\,s$^{-1}$), but we can still gain valuable insights into the nebular dynamics of NGC\,3242 from the 2D velocity maps. Based on the velocity map shown in Fig.~\ref{velocity}, k4 is blue-shifted, while k1 is red-shifted, with a relative velocity difference of 60~km\,s$^{-1}$ between them .

\subsection{Atomic gas and first detection of the far-red [C I] $\lambda$8727}

\begin{figure}[h!]
    \centering
    \includegraphics[width =0.45\textwidth]{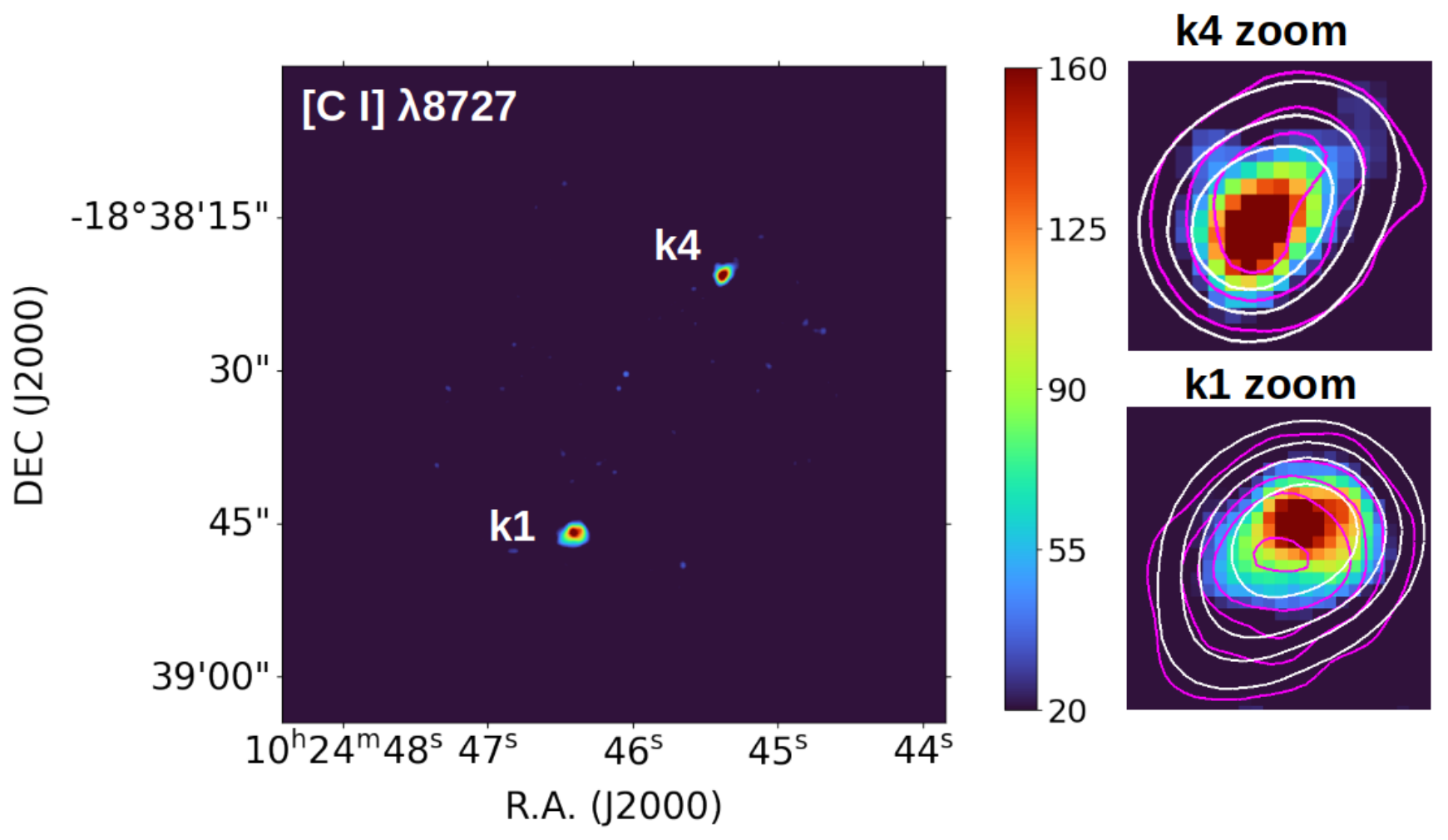}
    \caption{The detection of {[C~\sc i]} $\lambda$8727 at k1 and k4 LISs. The zoomed in figures show also the contours for the {[O~\sc i]} $\lambda$6300 (white) and {[N~\sc i]} $\lambda$$\lambda$5198,5200 (magenta) emission lines. The images are smoothed with $\sigma=1.5$. The color bar represents the values of the pixels (in flux units x10$^{-20}$ erg\,s$^{-1}$\,cm$^{-2}$\,spaxel$^{-1}$). }
    \label{CI_OI}
\end{figure}

Many studies have so far revealed the presence of atomic carbon lines such as [C~{\sc i}] $\lambda\lambda\lambda$8727, 9824, 9850 in PNe, \citep[see table 1 in][and references therein]{Akras_2024}. However, the majority of these studies were conducted using long-slit spectroscopy, making it unclear which specific nebular region predominantly emits these lines. In this context, the investigation of MUSE data for NGC\,3242 revealed the presence of the far-red atomic {[C~\sc i]}~$\lambda$8727 emission line, mainly emitted from its LISs, with a flux of around 2.5$\times$10$^{-16}$~erg\,s$^{-1}$\,cm$^{-2}$ in both sides. The left panel of Fig.~\ref{CI_OI} shows the {[C~\sc i]} $\lambda$8727 emission line map. Recently, the same far-red {[C~\sc i]} emission line was detected in the LISs of NGC\,7009 using MUSE data \citep{Akras_2024}. The detection  of {[C~\sc i]} emission in LISs, along with the enhanced emission of low-ionization lines such as {[N~\sc ii]}, {[S~\sc ii]}, {[O~\sc i]} supports the idea that LISs are potential mini-photo-dissociation regions (PDRs) embedded in PNe \citep{Aleman_2011}. After all, the {[C~\sc i]} $\lambda$8727 line has already been detected in PDRs within {H~\sc ii} regions, \citep[see][and references therein]{Hii_regions}.

\begin{figure}[h!]
    \centering
    \includegraphics[width =0.5\textwidth]{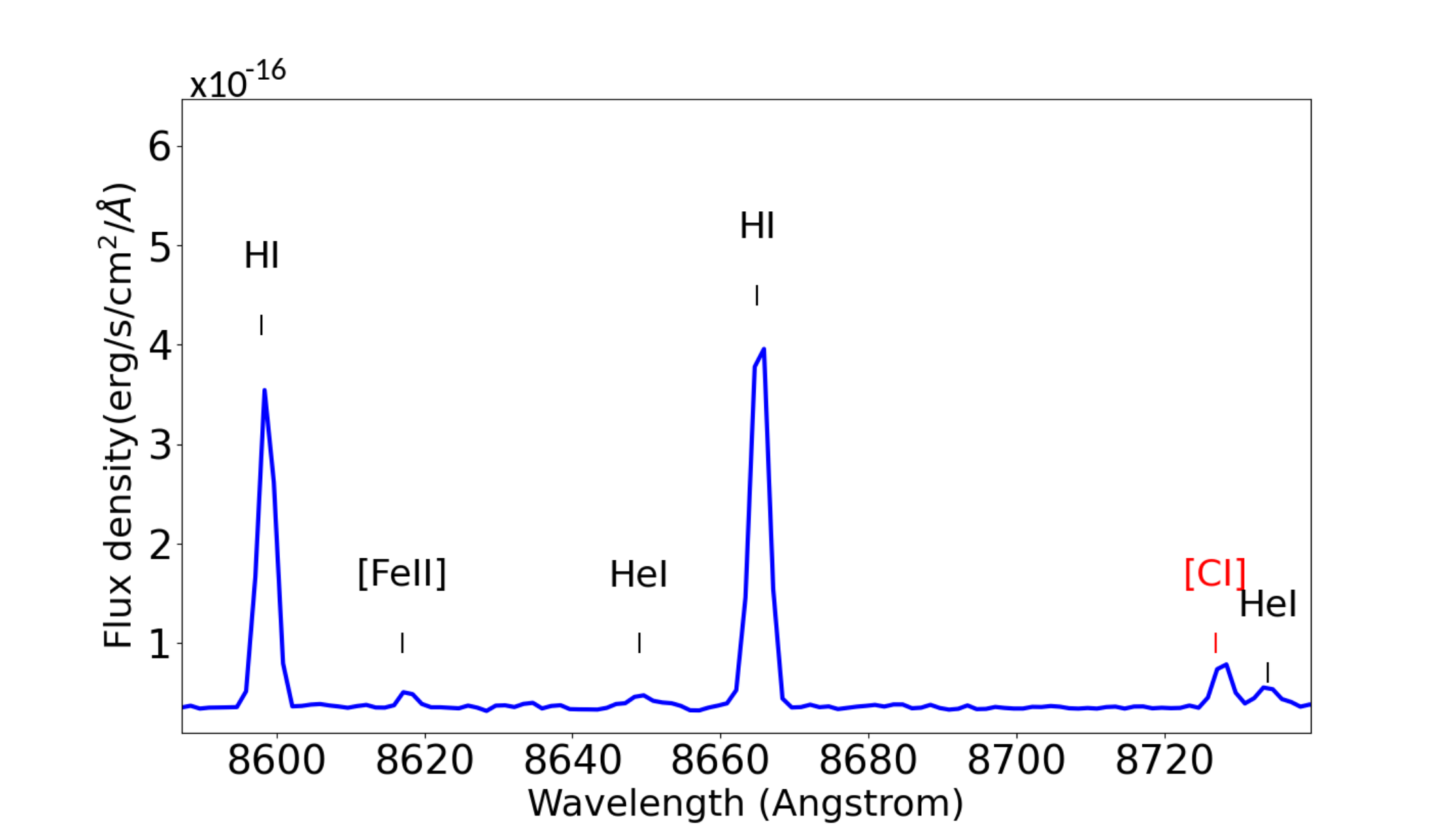}
    \caption{Extracted spectrum from MUSE datacube in the region of k1. The {[C~\sc i]} $\lambda$8727 emission line is marked with red color. The {[Fe~\sc ii]} $\lambda$8617 emission line is also identified in the spectrum.}
    \label{spec_CI}
\end{figure}

The atomic gas of the LISs is also characterized by the emission of {[O~\sc i]} and {[N~\sc i]} lines \citep[see theoretical work from][]{Aleman_2011}. Here, the {[N~\sc i]} $\lambda$$\lambda$5198, 5200 doublet was found to emanate from the LISs of NGC\,3242 with a mean flux, between k1 and k4, of 2.3$\times$10$^{-15}$ erg\,s$^{-1}$\,cm$^{-2}$, while {[O~\sc i]} $\lambda\lambda$6300, 6363 emission lines were detected with mean fluxes, between k1 and k4, 4.1$\times$10$^{-14}$ and 1.3$\times$10$^{-14}$ erg\,s$^{-1}$\,cm$^{-2}$, respectively. The line fluxes for the aforementioned atomic lines were estimated in a region of 5$^{\prime\prime}\times$4$^{\prime\prime}$ centered at k1 or k4 (see Fig.~\ref{emission_maps}) The contours of {[O~\sc i]} $\lambda$6300 and {[N~\sc i]} $\lambda\lambda$5198, 5200 emission lines on the {[C~\sc i]} $\lambda$8727 map
are presented in the right panels of Fig.~\ref{CI_OI}. An overlap is observed between {[O~\sc i]} and {[N~\sc i]} while there is small offset with the emission of {[C~\sc i]} (better seen in the radial profiles presented in Sect. \ref{radialprof_sect})

\subsection{Iron emission in NGC\,3242}
\label{Iron_emission}
\begin{figure}[h!]
    \centering
    \includegraphics[width =0.44\textwidth]{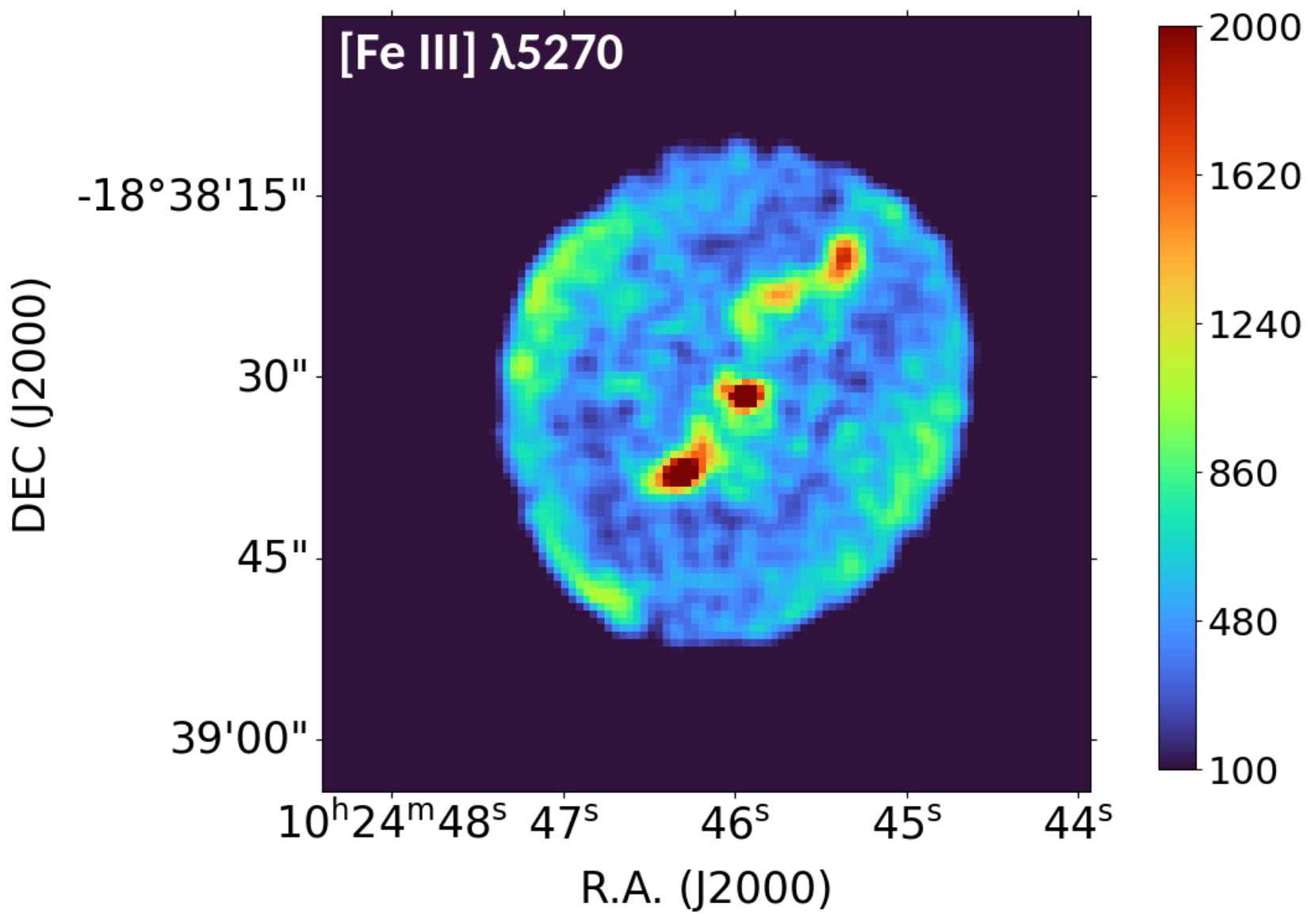}
    \caption{[Fe~{\sc iii}] $\lambda$5270 emission line map. The map is scaled down 3 times and smoothed with a $\sigma$=1. The color bar represents the values of the pixels (in flux units x10$^{-20}$ erg\,s$^{-1}$\,cm$^{-2}$\,spaxel$^{-1}$).}
    \label{FeIII_map}
\end{figure}

A weak detection of {[Fe~\sc ii]} $\lambda$8617 emission line can be seen in the spectrum of the k1 LIS (Figure \ref{spec_CI}). Generally, {[Fe~\sc ii]} is considered as shock tracer and is usually detected in shock excited regions of gaseous nebulae such as SNRs \citep[e.g.,][]{Temim2024} or Herbig-Haro objects \citep[e.g.,][]{Reiter2024}. In the case of NGC\,3242, this emission is concentrated in the k1 and k4, which could indicate shock origin for the LISs.
A jet-like structure is also unveiled in the {[Fe~\sc iii]} $\lambda$5270 emission line map, beginning close to the central star and pointing towards the direction of the LISs. 
The jet appears asymmetric with the northwest part, being more extended than the southeast part (see Fig.~\ref{FeIII_map}), due to the orientation of the nebula \citep{Morphokinematcs}.

\subsection{Radial Profiles of typical emission lines}
\label{radialprof_sect}

The {\sc satellite} radial analysis module was used to examine how the surface brightness of typical emission lines changes with the distance from the CS, and to detect the ionization stratification at the position of the LISs. Two different pseudo-slits were used, placed at position angles of 160\degr~and 320\degr~(from the north and counterclockwise). In each case, the pseudo-slit had dimensions 19$^{\prime\prime}$~$\times$~2$^{\prime\prime}$, extending from the CS to the outer regions of the nebula. Knot k1 is located 13$^{\prime\prime}$ from the CS, while k4 is at a distance of 16$^{\prime\prime}$.

\begin{figure}[h!]
    \centering
    \null\hfill
    \subfloat{\includegraphics[width =0.5\textwidth]{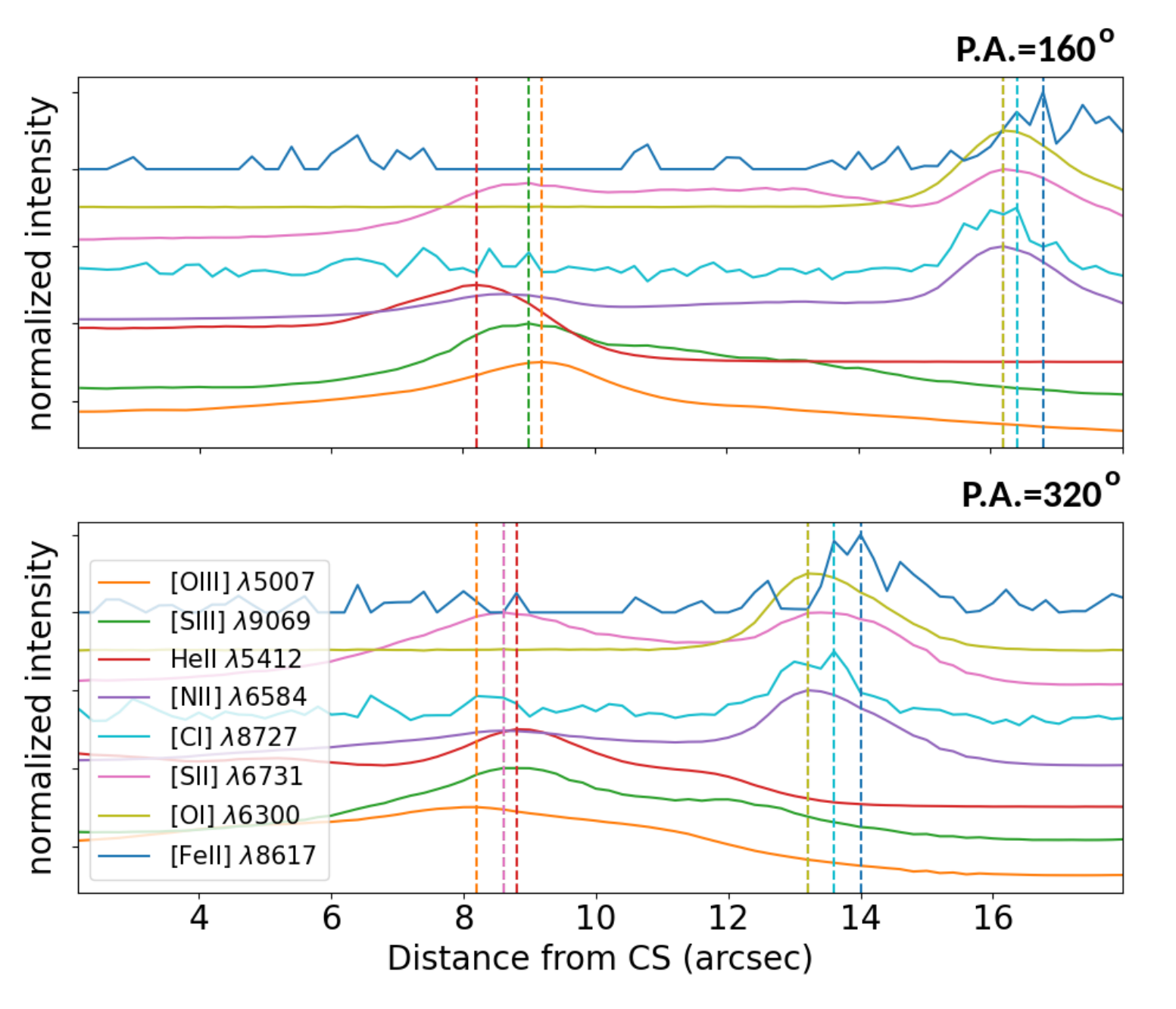}}
    \caption{The radial profiles of some typical optical emission lines for P.A. = 160\degr~(top panel) and P.A. = 320\degr~(bottom panel). The line intensities are normalized and an offset of 0.5 has been applied between the lines. The dashed lines define the pick intensities.   There is an overlap in the picks of {[N~\sc ii]}, {[S~\sc ii]} and {[C~\sc i]} for both P.A. and between {[S~\sc iii]} and the first pick of {[S~\sc ii]} only for P.A.=360\degr. }
    \hfill
    \label{rad_prof}
\end{figure}

The intensities of moderate and high ionization lines of {He~\sc ii}, {[O~\sc iii]} and {[S~\sc iii]} show peaks at distances closer to the CS and near the rim. In contrast, the low-ionization lines {[S~\sc ii]}, {[N~\sc ii]},  {[Fe~\sc ii]} and the atomic lines of {[C~\sc i]} and {[O~\sc i]} show a prominent peak at the position of LISs (see Fig.~\ref{rad_prof}). {[S~\sc ii]} exhibits a first pick at the rim too. This is reasonable, since S$^{+}$ requires photon energies $\sim$ 10.4~eV, which is lower than the ionization potential of N$^{+}$ ($\sim$ 14.5~eV). So, S$^{+}$ is created while N$^{+}$ is still shielded by hydrogen, leading to the emission from different regions. In general, no significant spatial offset is found between the low-ionization lines. In particular, all lines exhibit a peak at (16.0-16.2)$^{\prime\prime}$ and (13.2-13.6)$^{\prime\prime}$ along the direction of k1 and k4, respectively. Surprisingly, an offset of $\sim$0.5$^{\prime\prime}$  is detected in {[Fe~\sc ii]} compared to {[C~\sc i]} in both pseudo-slit directions (further analysis on this issue in Bouvis et al. in prep.). Note that MUSE's spatial sampling is 0.2$^{\prime\prime}$. Moreover, our results align well with the ionization stratification found in the LISs of NGC\,7009 by \cite{LIS_H2_Akras, Concalves_2003}.

On the contrary, the peak emission of moderate/high ionization lines exhibit different trend in both directions. Along the direction of k1 (P.A. of 160 degrees), {He~\sc ii} line peaks first at 8.2$^{\prime\prime}$ from the CS and then the {[O~\sc iii]} and {[S~\sc iii]} lines peak further away, at $\sim$9.2$^{\prime\prime}$. On the other hand, in the direction of k4 (P.A.~320 degrees) {[O~\sc iii]} peaks at 8.2$^{\prime\prime}$ and then {[S~\sc iii]} and {He~\sc ii} follow at distances of
8.6$^{\prime\prime}$ and 8.8$^{\prime\prime}$ from the CS, respectively.

\subsection{Nebular shell}

Through a detailed investigation of the NGC\,3242 MUSE data, new structures were discovered perpendicular to the pair of LISs, extending from the nebular rim to the inner shell. The enhanced emission is mainly observed in the {[S~\sc iii]} and {[N~\sc ii]} emission line maps (see bottom panels in Fig.~\ref{emission_maps}). Two faint arc-like structures are also visible above k1 and k4 at distances of 21$^{\prime\prime}$ and 19$^{\prime\prime}$ from the CS, respectively. The two arcs are radially symmetric with respect to the CS and are also aligned with k4 and k2, while an offset is observed in the case of k1 (see Fig.~\ref{emission_maps}).

Six pseudo-slits were placed along the identified structures, shown as blue rectangles in the bottom panels of Fig.~\ref{emission_maps}. The six regions defined by these pseudo-slits did not show significant differences among them. The mean values for their physical properties and some emission line ratios are presented in the second column of Table~\ref{table_shell_knots}.
The bright structures at the leading edges of the nebular inner shell are more prominent in {[\sc S~iii]} and {[\sc N ii]}. To our knowledge, this is the first time that these structures are detected in NGC\,3242. Moreover, {[Fe~\sc iii]} emission has also been detected from the same structures (see Fig.~\ref{FeIII_map}). 
To assess if these structures could result from shock interactions, the logarithmic ratios of {[\sc S ii]}/H$\rm\alpha$ and {[N~\sc ii]}/H$\rm\alpha$ were examined \citep{Sabbadin_diagnostics, diagnostics, leonidaki}. We estimated mean values of log({[\sc S ii]} (6316+6731)/H$\rm\alpha$) = $-$2.54$\pm$0.03 and log({[N~\sc ii]} (6548+6584)/H$\rm\alpha$) = $-$2.04$\pm$0.03, indicating that photoionization is likely the prevailing mechanism in these regions \citep{diagnostics_Akras}. The electron temperature in these regions is $T_{\rm e}$~({[S~\sc iii]}) = (11\,600$\pm$660)~K, while the electron density is $n_{\rm e}$~({[Cl~\sc iii]}) = (900$\pm$500)~cm$^{-3}$. Both $T_{\rm e}$ and $n_{\rm e}$ are consistent with the values estimated for the gas around the new structures (blue regions in the bottom panels of Fig.~\ref{emission_maps}, see last column of Table \ref{table_shell_knots}).

\subsection{Electron temperature and electron density from CELs}

The extinction coefficient c(H$\rm\beta$), the electron temperature ($T_{\rm e}$) and density ($n_{\rm e}$), are key parameters for the physical characterization of a nebula.
Here, the rotational analysis module is employed to test the behavior of c(H$\rm\beta$), $T_{\rm e}$ and $n_{\rm e}$ with the position angle of a pseudo-slit. For this purpose, we considered pseudo-slits extending from the CS to the outer nebular regions, with the P.A. increasing in steps of 10\degr~from 0\degr~to 360\degr. 
In Fig.~\ref{knots_sp_slit}, we show the behavior of c(H$\rm\beta$) (upper panel), $T_{\rm e}$ ({[\sc S iii]}) (middle panel), and $n_{\rm e}$ ({[S~\sc ii]}) (lower panel), computed from the different pseudo-slits described above. 
We found that c(H$\rm\beta$) is almost constant throughout the nebula, with a mean value of 0.140$\pm$0.024. On the other hand, $T_{\rm e}$ ({[\sc S iii]}) also remains almost constant (between 14\,000 K and 15\,000 K) for most P.A. values, with a significant decrease of about 4\,000 K around P.A. 160\degr~and 320\degr, which coincides with the orientations of the LISs. Furthermore, $n_{\rm e}$ ({[S~\sc ii]}) is around 2\,000 cm$^{-3}$ for most P.A., but reaches higher values also in the direction of the LISs. It is important to note that these values do not necessarily imply an actual increase in $n_{\rm e}$ (or decrease in $T_{\rm e}$) at the LISs, since the resulting $T_{\rm e}$ and $n_{\rm e}$ are the integrated values along the pseudo-slits. It is worth noticing, though, that from the radial analysis, it is evident that the position of the slit can yield to notable discrepancies in $T_{\rm e}$ and $n_{\rm e}$ values. In the following paragraphs, we examine the 2D spatial distribution of both $T_{\rm e}$ and $n_{\rm e}$ which provides further insight into what is occurring at the LISs. 

\begin{figure*}[h!]
    \centering{
    \null\hfill
    \caption*{}
    \hfill
    \subfloat{\includegraphics[width =0.7\textwidth]{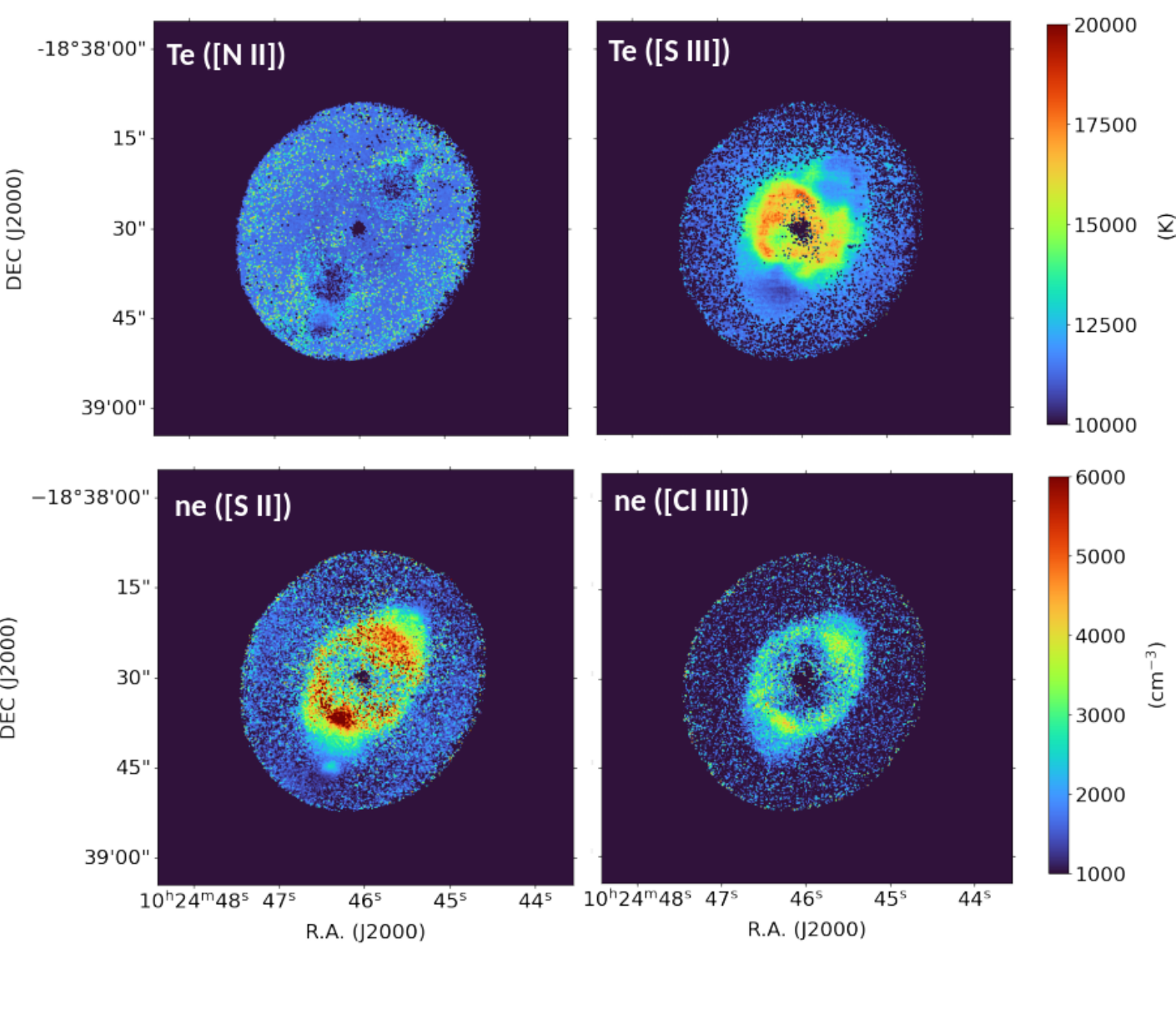}}
    \null\hfill
    }
\caption{Electron temperature and electron density 2D maps from CELs. $T_{\rm e}$ is calculated from {[N~\sc ii]} and {[S~\sc iii]} emission maps (top panel left and right respectively). $n_{\rm e}$ is calculated from {[S~\sc ii]} and {[Cl~\sc iii]} emission maps, assuming $T_{\rm e}$ ([N{~\sc ii}]) and $T_{\rm e}$ ({[S~\sc iii]}), respectively (bottom panel left and right, respectively).}
\label{TeNe}
\end{figure*}

In the {$T_{\rm e}$ ([S~{\sc iii}]) 2D map, presented in the upper right panel of Fig.~\ref{TeNe}, it is evident that $T_{\rm e}$ is higher at the nebular rim, where it reaches values of $\sim$17\,000 K. This behavior has also been previously observed in {$T_{\rm e}$ ([O~{\sc iii}]) \citep{FLIERS, Monteiro2013}. The NW and SE parts of the rim are not as extended, and they do not significantly affect the integrated value of the electron temperature found through the rotational analysis. In contrast, along the minor nebular axis, the contribution of the rim to $T_{\rm e}$ is not negligible, resulting in higher values when the slit analysis is considered. The mean $T_{\rm e}$ {([S~\sc iii])} from the 2D map is (12\,500$\pm$1\,700) K.

\begin{figure}[h!]
     \centering
	\begin{subfigure}{2.0\linewidth}
	    \includegraphics[width =0.5\textwidth]{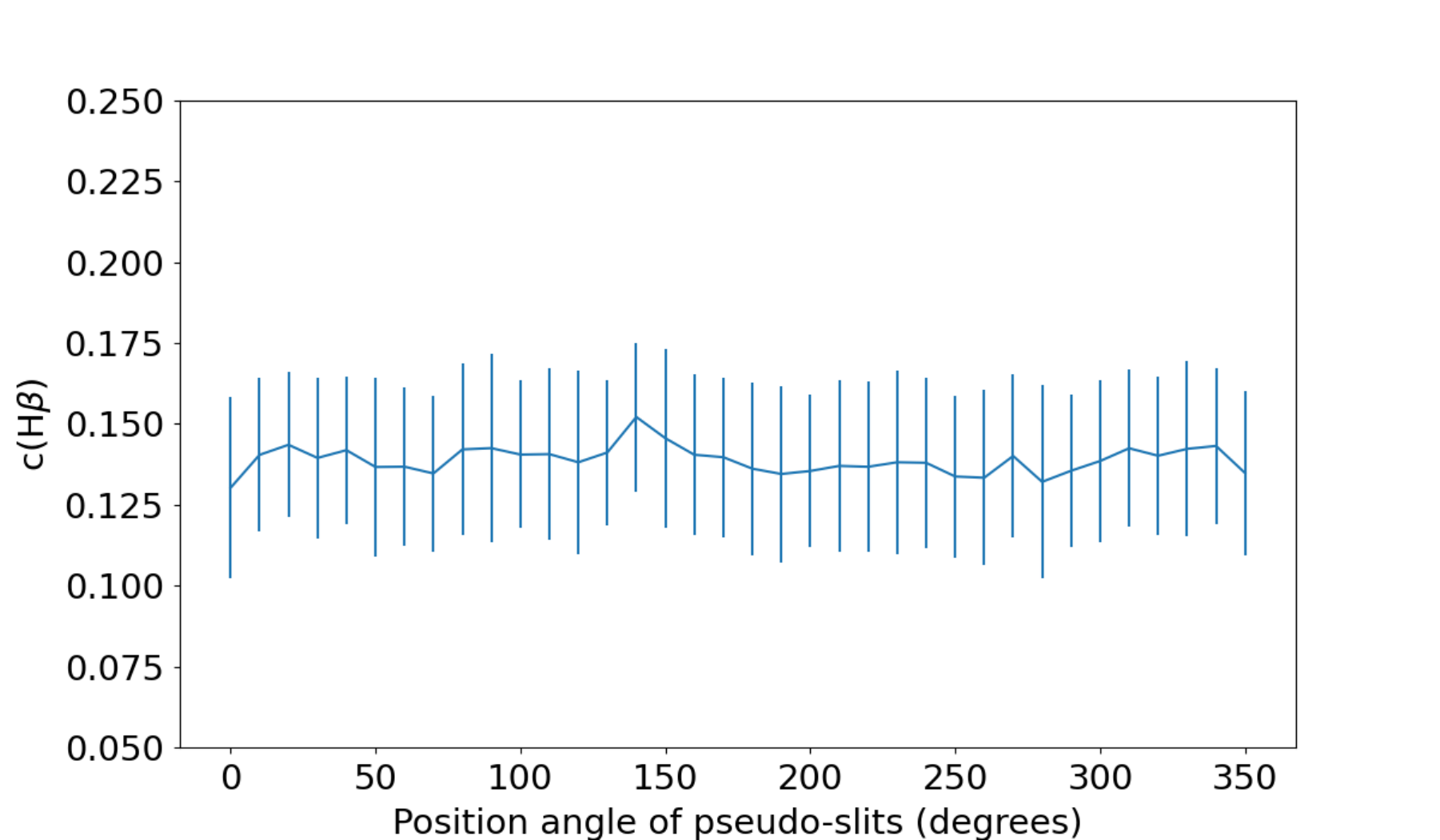}\\
	    \includegraphics[width =0.5\textwidth]      {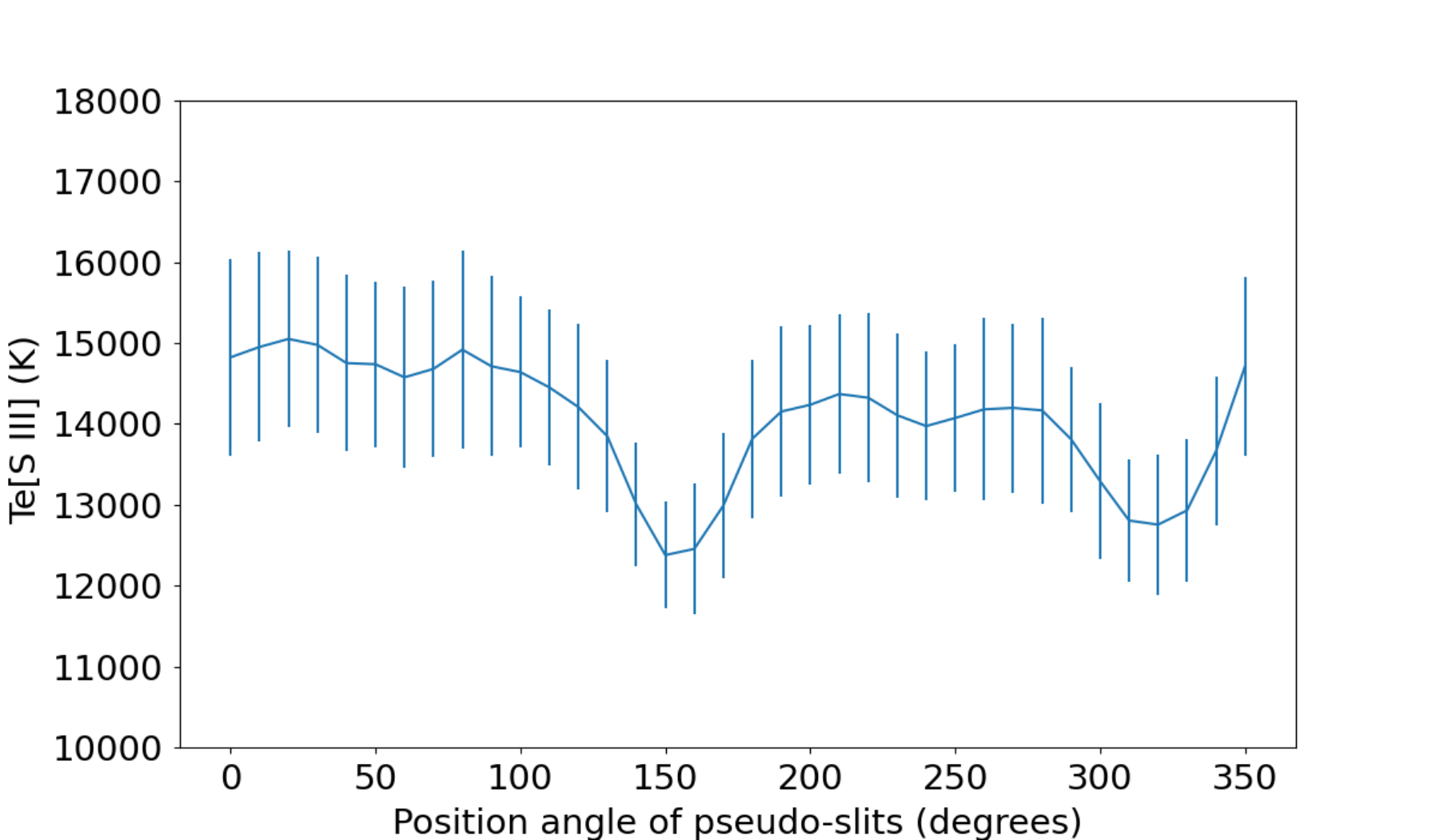}\\
            \includegraphics[width =0.5\textwidth]{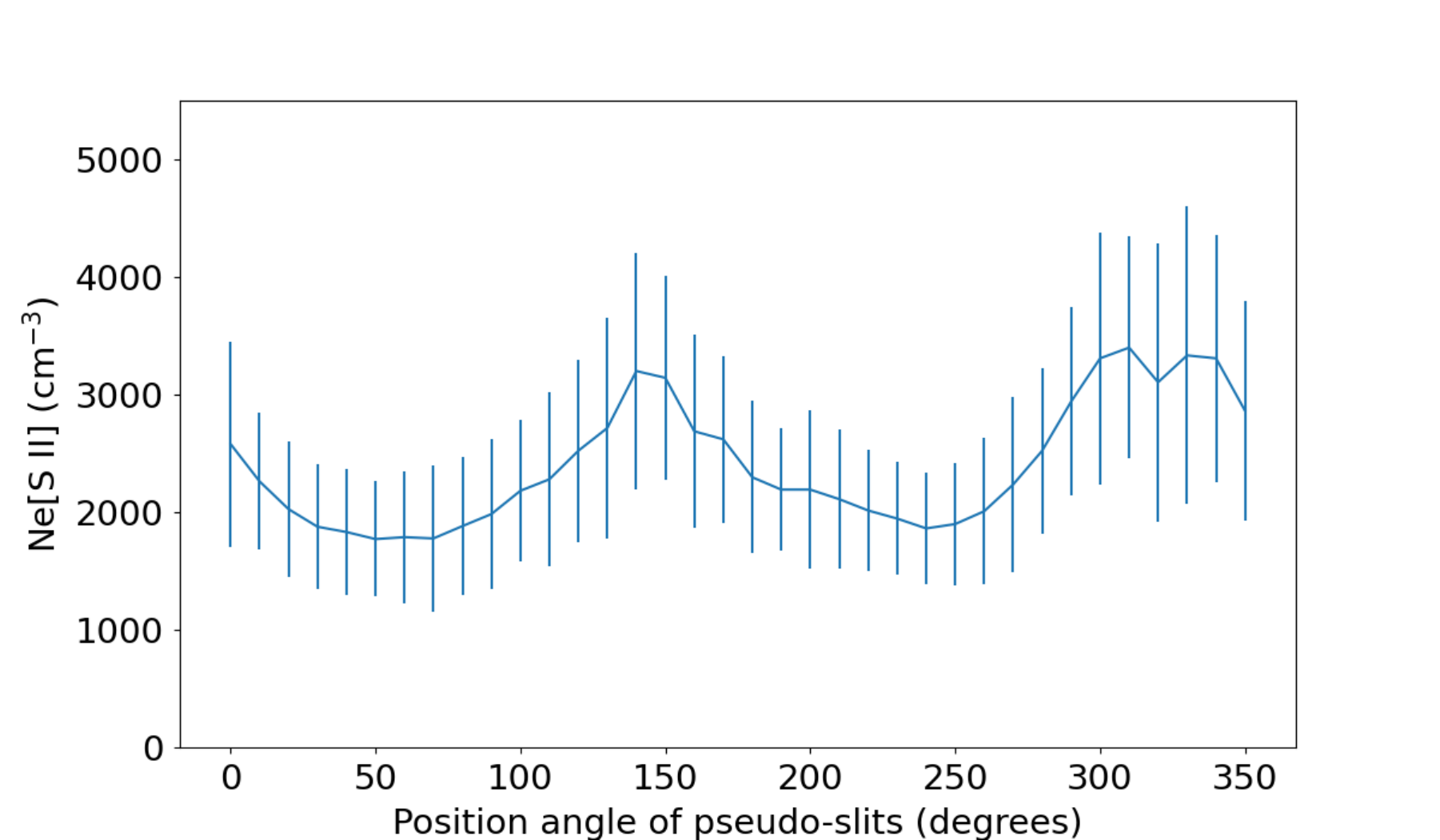}
            
	\end{subfigure}
     \caption{From the top to the bottom, they are c(H$\rm\beta$), {$T_{\rm e}$\sc[S iii]} and {$n_{\rm e}$\sc[S ii]} versus the P.A. of the pseudo-slit from 0\degr~to 360\degr. The error-bars are estimated from 100 Monte Carlo iterations.}
     \label{knots_sp_slit}
\end{figure}

In the upper-left panel of Fig.~\ref{TeNe}, we also present the electron temperature calculated from {\sc [N ii]} diagnostic lines. It is important to note that, due to the mask that we applied to the {\sc [N ii]} $\lambda$5755 line map (see Sect.~\ref{sec:masks}), {$T_{\rm e}$~([N~\sc~ii}]) is only well-defined in the unmasked regions of the nebula, which correspond to the lower ionization degree zones around the LISs. In this case, $T_{\rm e}$ {\sc([N ii])} has a mean value of (11\,700$\pm$1\,500) K, in good agreement with previous studies \citep{Krabbe, Pottasch, Monteiro2013}.

As for the electron density, we calculated a mean {$n_{\rm e}$ ([S~\sc ii])} of (2\,200$\pm$1\,500)~cm$^{-3}$ and {$n_{\rm e}$~([Cl~\sc iii])} of (1\,600$\pm$1\,000)~cm$^{-3}$ from the 2D density maps (lower panels of Fig.~\ref{TeNe}), given {$T_{\rm e}$ ([N~{\sc ii}]) and {$T_{\rm e}$ ([S~{\sc iii}]), respectively. In both cases, the electron density of the LISs is lower than the surrounding nebula. The fact that LISs are less dense than the surrounding nebula is in conflict with the theoretical scenarios about their formation \citep[e.g.,][]{Steffen_2001, Raga_2008}. 
On the contrary, the highest values for $n_{\rm e}$ are observed at the nebular rim, which explains the increase that was previously noticed at P.A. 160\degr~and 320\degr~through the rotational analysis. The mean values, the median, the Q3 and 95\% percentiles and the standard deviation for c(H$\rm\beta$), $T_{\rm e}$ and $n_{\rm e}$ are also listed in Table \ref{table_Te_Ne}.

\begin{table}[h!]
    \centering
     \caption{The median values, the 75\%\ percentile (Q3), the 95\% percentile, the mean and the standard deviation ($\sigma$) are presented for c(H$\rm\beta$), $T_{\rm e}$ in K and $n_{\rm e}$ in cm$^{-3}$ of the entire PN}
    \label{table_Te_Ne}
    \begin{tabular}{|p{1.6cm}ccccc|}
    \hline
    & median& Q3 & 95\% &mean & $\sigma$ \\
     \hline
     \hline
      c(H$\rm\beta$) &0.14& 0.18& 0.24 & 0.14& 0.06\\ 
    $T_{\rm e}$ \textsc{([N ii])}  &11\,300 & 11\,400 & 15\,700 &  11\,700& 1\,500\\ 
    $T_{\rm e}$ \textsc{([S iii])}  & 11\,900 & 12\,900 & 16\,400& 12\,500&1\,700\\ 
    $T_{\rm e}$ (He~\sc i)  & 5\,400 & 4\,000& 7\,200 & 5\,700 &2\,200 \\ 
      & 7\,800$^\dag$ & 8\,300$^\dag$& 9\,100$^\dag$ & 7\,800$^\dag$&1\,000$^\dag$\\ 
    $T_{\rm e}$ (PJ)  & 8\,000  & 11\,400 &  20\,700 &9\,300 &5\,300 \\
     & 9\,000$^\dag$ &10\,200$^\dag$ & 15\,800$^\dag$ &8\,900$^\dag$& 3\,700$^\dag$\\
    $n_{\rm e}$ \textsc{([S ii])}& 1\,700 & 2\,600& 4\,900& 2\,200 &1\,500\\ 
    $n_{\rm e}$ ([{Cl~\sc iii}]) & 1\,400 & 2\,200 & 3\,400 & 1\,600 & 1\,000\\  
    \hline
    \end{tabular}
    \tablefoot{
    \tablefoottext{$\dag$} {$T_{\rm e}$ (PJ), $T_{\rm e}$ (He~{\sc i}) and $\sigma$, estimated in the elliptical annulus of Fig.~\ref{TeORL}.\\ }
    }

\end{table}

\subsection{Electron temperature from ORLs}

Certain recombination lines can also be used for the estimation of $T_{\rm e}$. Based on He~{\sc i} 7281/6678 diagnostic ratio,  $T_{\rm e}$ can be estimated either by an analytic formula or by a linear representation. The latter approach is only valid in the low $T_{\rm e}$ regime ($T_{\rm e}$<15\,000 K). The analytic formula involves the corrected intensities of He~{\sc i} $\lambda\lambda$7281, 6678 lines and three fitting parameters: a$_i$, b$_i$ and c$_i$ \citep[see relation (4) from][]{Zhang2005nonlinear}, which are associated with the line emissivities. The fitting parameters were estimated following the procedure described by \cite{Benjamin1999}, but using the most recent emissivities from \citep{Porter_2012,Porter_2013} available in {\sc PyNeb} \citep{pyneb, pyneb2_2020}. These fitting parameters are also density dependent, so a$_i$, b$_i$, and c$_i$ were first estimated for $n_{\rm e}$ values of 100, 1,000, and 10,000 cm$^{-3}$, and then a minimization process was applied to refine them, given the $n_{\rm e}$ computed for NGC\,3242.

In cases of $T_{\rm e}$<15\,000 K, a linear representation is sufficient to estimate $T_{\rm e}$ {(He~\sc i)}. \cite{Zhang2005linear}, provided a linear representation for $T_{\rm e}$  with uncertainties up to 7\% and without dependencies on $n_{\rm e}$. Similarly, \cite{MendezDelgado2021}, provided another linear representation that varies with $n_{\rm e}$ at low densities, but remains constant at higher $n_{\rm e}$ values. This relationship involves the fitting parameters $\alpha$ and $\beta$, which were estimated from a linear interpolation of $n_{\rm e}$ values of NGC\,3242 over the range of $n_{\rm e}$ provided in table~D7 of \cite{MendezDelgado2021}.
We applied each of these approaches and found that \cite{Zhang2005nonlinear} and \cite{MendezDelgado2021} are in good agreement for our case, with mean values $\sim$5\,650 K and $\sim$6\,000 K, respectively. In contrast, the \cite{Zhang2005linear} method tends to yield lower electron temperatures. The right panel of Fig.~\ref{TeORL} shows an example of $T_{\rm e}$ (He~{\sc i}) map generated using the approach of \cite{Zhang2005nonlinear}. Spatially, there is no significant variation among the inner nebular structures, although the rim exhibits slightly higher temperature. The lowest $T_{\rm e}$ ({He~\sc i}) is observed at the nebular shell, where it reaches values around 4\,000 K. The mean value of $T_{\rm e}$ ({He~\sc i}) is 5\,700~K (\ref{table_Te_Ne}), which is in good agreement with previous studies \citep{Tsamis}. It is also worth noticing that, the {He~\sc i} $\lambda$7281 emission line is fainter in comparison with the {He~\sc i} $\lambda$6678 emission, affecting $T_{\rm e}$ ({He~\sc i}) and leading to lower values at the nebular outer shell. In a very recent paper, \citet{MendezDelgado2024} has claimed that the lower $T_{\rm e}$ ({He~\sc i}) observed in photoionized nebulae relative to $T_{\rm e}$ ([{O~\sc iii}]) can be explained by a combination of photon loss from He$^+$ $n^1P \rightarrow 1^1S$ transitions (deviations from ``case B'') and the effect of temperature inhomogeneities in the nebula. We will discuss this further in Sect.~\ref{sec:discuss_phys}.

\begin{figure*}[h!]
    \centering{
    \null\hfill
    \caption*{}
    \hfill
    \subfloat{\includegraphics[width =0.8\textwidth]{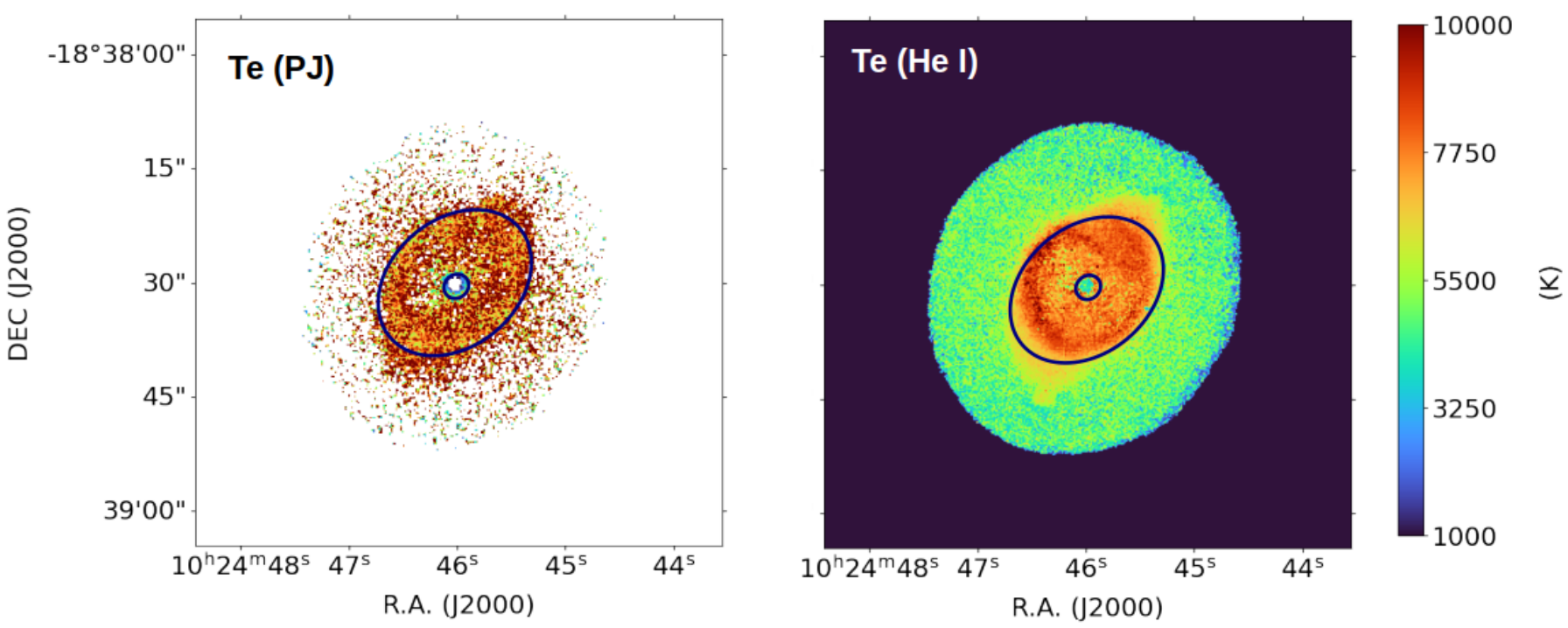}}
    \null\hfill
    }
\caption{Electron temperature maps from the ratio of the Paschen discontinuity to H~{\sc i} P9 line ($T_{\rm e}$(PJ), left panel), and from the He I $\lambda$7281/$\lambda$6678 recombination line ratio (right panel). The blue ellipse marks the boundaries of the inner shell.}
\label{TeORL}
\end{figure*}

The electron temperature can also be calculated from the ratio of the  H~{\sc i} continuum Balmer jump (BJ) or Paschen jump (PJ) at 3\,645~\AA~and 8\,200~\AA, respectively, to a nearby H~{\sc i} line. In the MUSE wavelength range, only the Paschen jump is available. For the estimation of the continuum before and after the PJ, the Continuum class of {\sc PyNeb} \citep{pyneb, pyneb2_2020} was used, assuming constant $n_{\rm e}$ =1.5$\times$10$^3$~cm$^{-3}$,  He$^{+}$/H$^{+}$~= 7.35$\times$10$^{-2}$ and He$^{2+}$/H$^{+}$ = 2.24$\times$10$^{-2}$. The continuum before and after the PJ was estimated at $\sim$8\,100~\AA~and $\sim$8\,400~\AA, respectively, normalized to H~{\sc i} P9 emission line at 9\,229 \AA. This line was chosen instead of P11, which is closer to the jump, because P11 is fainter, and we wanted to avoid the additional noise to the  $T_{\rm e}$ (PJ) map. Then we interpolated over the observed PJ, which was estimated by extracting continuum slices from the MUSE datacube, and normalizing them to the P9 emission line map. The 2D map that was constructed from this analysis is presented in the left panel of Fig.~\ref{TeORL}. At the outer nebular region, the temperature map is quite noisy and highly uncertain, probably due to the noisy continuum emission. The mean value is $\sim$9\,300~K, which is in good agreement with the value that was previously estimated by \cite{Tsamis} for NGC\,3242. Table \ref{table_Te_Ne}, shows that the median $T_{\rm e}$ ({He~\sc i}) is around 4\,500 K lower than $T_{\rm e}$ (PJ), in very good agreement with what was observed by \citet{Zhang2005nonlinear} in a sample of PNe, where they found an average difference of $T_{\rm e}$(H~{\sc i})$-$$T_{\rm e}$(He~{\sc i}) of approximately 4\,000~K. 
However, since both temperatures are not well defined outside the borders of the rim, the magnitude of this difference is weighted to the values obtained at the outer nebular regions. So, in order to properly compare them, we compared the median values in an ellipse containing the inner shell, but not the position of the central source (see Fig.~\ref{TeORL}). In this region, the median values of $T_{\rm e}$ ({He~\sc i}) and $T_{\rm e}$ (PJ) are (7\,800$\pm$1\,000) K and (8\,900$\pm$3\,700) K, respectively, which are now in better agreement than those estimated for the entire PN. We will discuss this issue further in Sect.~\ref{sec:discuss_phys}.

\subsection{Ionic and total abundances}
\label{Abundances_section}

Theoretically, the elemental abundance of a given element is simply the sum of the abundances of all its ionic species present in the gas. However, this is not the case when only the optical range is studied, and ICF must be applied to account for the unobserved ions. In this study, we adopt the ICF approach proposed by \cite{KB}. {\sc satellite} also estimates the ICFs recommended by \cite{DIMS}, but since NGC\,3242 is a high-excitation nebula, the criteria to estimate the ICFs from \cite{DIMS} were not satisfied in many cases. Using the specific slit analysis module of {\sc satellite} and selecting a pseudo-slit that covers the entire nebula (centered at the center of the PN with dimensions 40$^{\prime\prime}$$\times$50$^{\prime\prime}$), we performed an analysis of the integrated spectrum of NGC\,3242. This allowed us to calculate the physical conditions and the ionic and elemental abundances, which were then compared to the results from previous studies. In Figs.~\ref{abund_1} and ~\ref{abund_2} we present 2D ionic abundance maps for some species, along with the elemental abundance of He, which is the only element that does not require the use of an ICF. For the low ionization ions (e.g. N$^+$, S$^+$), the low excitation electron temperature $T_{\rm e}$~({[N~\sc ii]}) and density $n_{\rm e}$~({[S~\sc ii]}) were used to compute ionic abundances. For high ionization ions, such as O$^{2+}$, Cl$^{2+}$ or S$^{2+}$, $T_{\rm e}$ ({[S~\sc iii]}) and $n_{\rm e}$ ({[Cl~\sc iii]}) were employed. The second column of Table~\ref{table_abund} presents the computed physical conditions, as well as the ionic/total abundances from the specific slit analysis. 

In terms of spatial distribution, He/H does not vary significantly throughout the PN, but there is a trend toward higher values on the east side of the major axis (see Figs.~\ref{abund_1} and ~\ref{abund_2}). 
Regarding He ions, He$^{2+}$/H$^{+}$ is mainly concentrated in the inner nebular structures due to its higher ionization state, while He$^{+}$/H$^{+}$ is widely distributed throughout the PN.
N$^{+}$/H$^{+}$, O$^{+}$/H$^{+}$ and S$^{+}$/H$^{+}$ are more abundant in the LISs, as expected (Figs.~\ref{abund_1} and ~\ref{abund_2}). In contrast, O$^{2+}$/H$^{+}$ exhibits high values across the nebular structures, which corresponds to strong {[O~\sc iii]} emission, e.g. log({[O~\sc iii]} (4959+5007)/H$\rm\beta$) = 1.2.

\subsubsection{Effect of possible recombination contribution to the chemical abundances}
\label{recombination contrib}

At this point, it is worth mentioning that both {[N~\sc ii]} $\lambda$5755 auroral and {[O~\sc ii]} $\lambda$$\lambda$7320,7330 transauroral lines are potentially affected by recombination emission \citep{Liu2000}.

To correct the intensity of {[N~\sc ii]} $\lambda$5755 auroral line from this effect, we used two different approaches. First, we use the recipe provided by \cite{Liu2000} in their eq. 1. This approach is valid in the 5\,000 K $\leq$ $T_{\rm e}$ $\leq$ 20\,000 K range, which is what we expect in PNe with low abundance discrepancy factors (ADF), as is the case for NGC\,3242. Adopting $T_{\rm e}$ = 12\,700 K, and the N$^{2+}$/H$^{+}$ obtained from the {N~\sc ii} $\lambda$5679 recombination line, the recombination contribution is estimated to be $\sim$30\% for the integrated intensity of {[N~\sc ii]} $\lambda$5755 (normalized to H$\beta$=100) in the entire PN (for pseudo-slit's dimensions see Table \ref{table_abund}), while at the nebular rim is higher ($\sim$40\%). This is reasonable, since most of the N {\sc ii} $\lambda$5679 emission comes from the rim (see Fig.~\ref{NII_CII}). This result agrees within the errors with the one of \cite{Tsamis} using the N$^{2+}$/H$^{+}$ from recombination lines.

Another way to account for the recombination contribution to the {[N~\sc ii]} $\lambda$5755 line was provided by \citet{ADF_recomb}: 
\begin{equation}
   I(5755)_{cor} = I(5755)-\frac{j_{5755}(T_{\rm e},n_{\rm e})}{j_{5679}(T_{\rm e},n_{\rm e})}\times I(5679)
\end{equation}
Where j$_{5755}$ and j$_{5679}$ are the recombination emissivities. This approach was adopted by \citet{ADF_recomb} to account for the recombination contribution in high-ADF PNe, where a cold plasma component with $T_{\rm e}$ much lower than 5\,000 K, emits the bulk of metal recombination lines and, hence, it is out of the validity range for the use of \citet{Liu2000}'s formula. The flux of the  {N~\sc ii} $\lambda$5679 recombination line was extracted from the MUSE datacube and integrated for the entire nebula, and $T_{\rm e}$ = 12\,700 K and $n_{\rm e}$ = 1\,600 cm$^{-3}$ were adopted for the estimation of the recombination emissivities. As in \citet{ADF_recomb}, the recombination coefficients to compute the emissivities were adopted from \citet{Fang2011} and \citet{Pequignot1991} for the $\lambda$5679 and $\lambda$5755 recombination emission, respectively. Following this approach, the recombination contribution for [{N~\sc ii}] $\lambda$5755 emission line is estimated to be $\sim$30\%\ of the observed intensity, almost identical to the correction obtained following the previous approach. These results seem to confirm the validity of the \citet{Liu2000}'s formula in the  5\,000~K $\leq$ $T_{\rm e}$ $\leq$ 20\,000 K range.

Moreover, an inspection of the spatial distribution of the N~{\sc ii} $\lambda$5679 or C~{\sc ii} $\lambda$6461 recombination lines in Fig.~\ref{NII_CII} shows that there is no recombination emission in the LISs, but only in zones internal to the rim.  At the nebular rim, the corrected intensity of {[N~\sc ii]} $\lambda$5755 is 0.05 instead of 0.08 (normalized to H$\beta$=100), while $T_{\rm e}$ ([N {\sc ii}]) is (10\,900$\pm$600)~K instead of (12\,900$\pm$760)~K, indicating a decrease of $\sim$2\,000 K.
The recombination correction was applied to the [N {\sc~ii}] $\lambda$5755 emission line map, since both N {\sc ii} $\lambda$5679  and [N {\sc~ii}] $\lambda$5755 emission lines are intense at the nebular rim. Nonetheless, when estimating the chemical abundances of the PN, considering the mask applied to [N {\sc~ii}] $\lambda$5755, the effect of the recombination contribution drops to $\sim$4\%. However, we accounted for this correction and re-estimated the ionic/total abundances of the nebula, which are listed in the first column of Table \ref{table_abund}.

\begin{figure}[h!]
     \centering
	\begin{subfigure}  {2.0\linewidth}
        \includegraphics[width =0.45\textwidth]{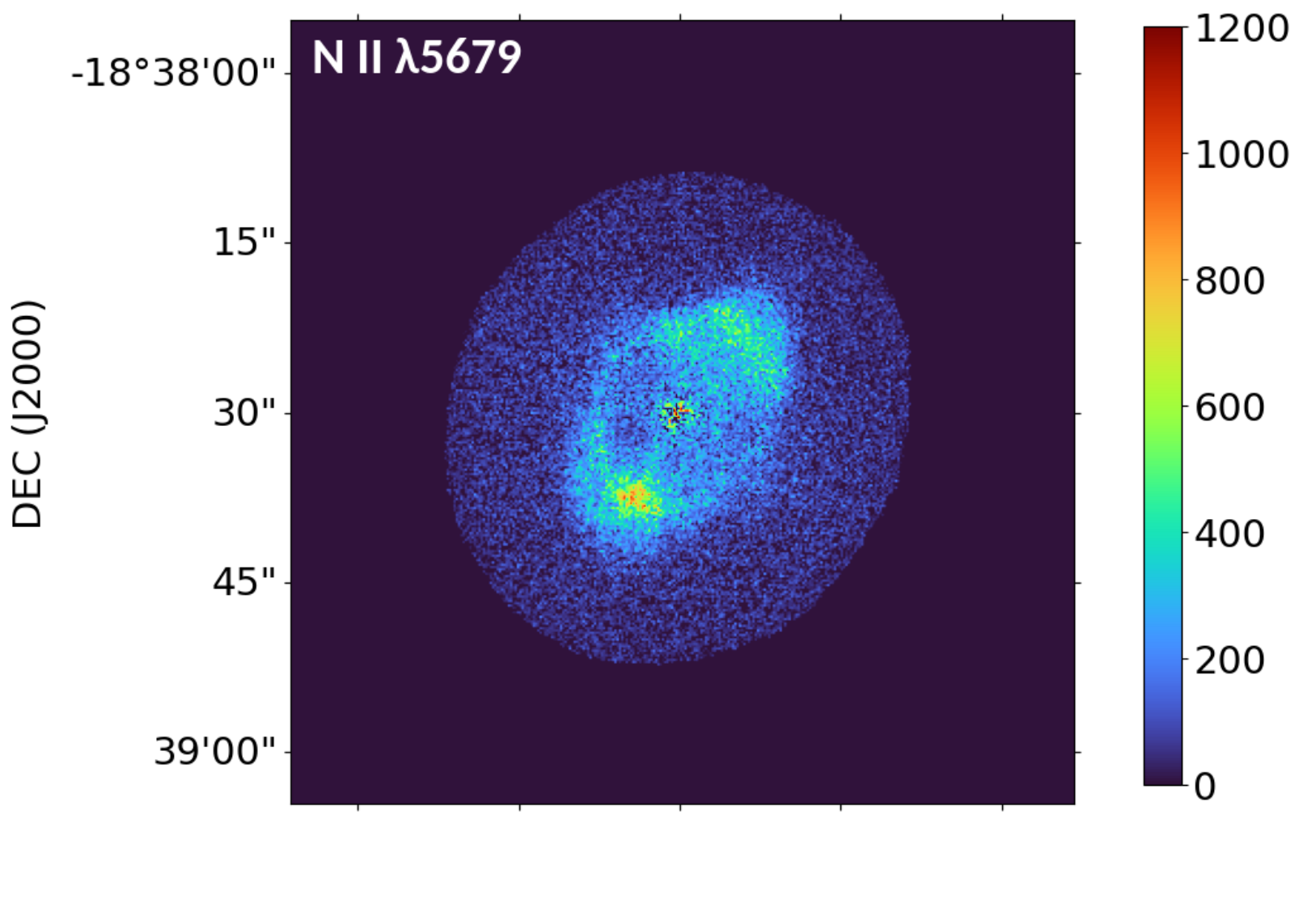}\\ 
        \includegraphics[width =0.45\textwidth]{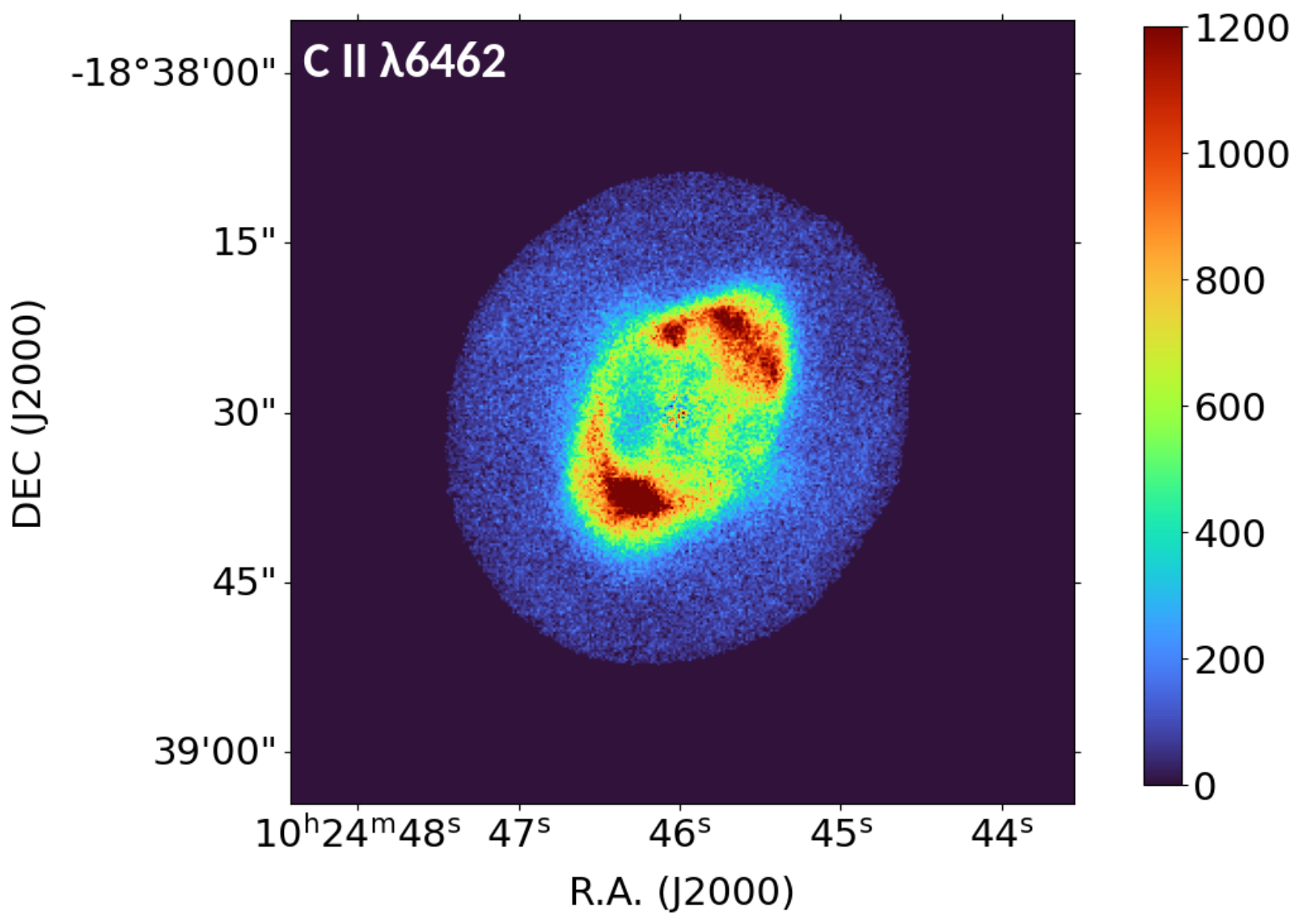}
        
	\end{subfigure}
     \caption{{N~\sc ii} $\lambda$5679 (top panel) and {C~\sc ii} $\lambda$6462 (bottom panel) emission line maps. The color bar represents the values of the pixels (in flux units x10$^{-20}$ erg\,s$^{-1}$\,cm$^{-2}$\,spaxel$^{-1}$).}
     \label{NII_CII}
\end{figure}

Regarding the {[O~\sc ii]} $\lambda$$\lambda$7320,7330 lines, we could not adopt the approach by \citet{ADF_recomb} because the nominal wavelength range used for our NGC\,3242 MUSE data does not cover wavelengths below 4800 \AA, skipping the detection of the {O~\sc ii} recombination lines at $\sim$4650 \AA. So, we employed the eq.~2 from \cite{Liu2000} to account for the recombination contribution.

As we only have [O~{\sc iii}] collisionally excited lines in our spectrum, we adopt O$^{2+}$/H$^{+}$ = 2.11x10$^{-4}$ and, as in the case of [N~{\sc ii}] $\lambda$5755, $T_{\rm e}$ =12\,700 K. With these values, the recombination contributed is estimated to be $\sim$20\%\ of the observed intensity relative to H$\beta$=100, which is in good agreement, within the errors, with previous studies \citep{Tsamis}. The corrected intensity is, therefore, $I$({[O~\sc ii]} $\lambda$$\lambda$7320,7330)/ $I$(H$\rm\beta$)=0.85 which leads to O$^{+}$/H$^{+}$ = 6.70$\times$10$^{-6}$ and O/H = 2.50$\times$10$^{-4}$, instead of 8.18$\times$10$^{-6}$ and 2.60x10$^{-4}$, respectively. However, it is worth mentioning that although the total O abundance is little affected, the ionization degree of the nebula, given by the O$^+$/O$^{2+}$ ratio, is strongly affected. On the other hand, if we adopt the O$^{2+}$/H$^+$ obtained from recombination lines by \citet{Tsamis}, we obtain a recombination contribution that is even higher to the measured intensity of the line. This is a weird result as most of the emission of the {[O~\sc ii]} $\lambda$$\lambda$7320,7330 lines arises from the LISs, where no recombination emission from heavy elements is seen (see Fig.~\ref{NII_CII}). \citet{GomezLlanos2024} found a similar overestimation of recombination contribution using this formula in their analysis of MUSE data of NGC\,6153. We, therefore, recommend adopting, whenever possible, both \citet{Liu2000} and \citet{ADF_recomb} approaches to check consistency between them. We also refer the reader to the discussions by \citet{Liu2000, ADF_recomb} and \citet{GomezLlanos2024} on this issue.
Due to the lack of information on both the total flux and the spatial distribution of O~{\sc ii} recombination lines, we decided not to apply such corrections to the emission maps of [O {\sc ii}].

\subsubsection{Comparison with previous studies}

In general, several studies have examined the chemical composition of NGC\,3242, using both IFU data \citep[][MG13]{Monteiro2013} and slit spectroscopy \citep[][MH16, PB08, KC06 and TS03, respectively]{Miller_2016, Pottasch, Krabbe, Tsamis}. MG13 and KC06 studied this PN using optical data, while MH16 used HST spectra ranging from ultraviolet (UV) to near-infrared (NIR), obtained during the HST Cycle 19 \citep{Dufor_2015}. In the study of PB08, UV, optical, and NIR data were used as well, while TS03 scanned several PNe using a slit in the optical spectrum, supplementing the analysis with archival UV data from the International Ultraviolet Explorer (IUE). Table~\ref{table_abund} presents the physical ($T_{\rm e}$, $n_{\rm e}$) and chemical properties of NGC\,3242 from a pseudo-slit that covers the entire nebula (described in Sect. \ref{Abundances_section}), as well as the results from previous studies. All the studies mentioned above used ICF schemes to compute elemental abundances, except PB08, who took advantage of a multiwavelength analysis and considered ICF = 1 for every element.

Regarding the physical parameters, $T_{\rm e}$ and $n_{\rm e}$, the estimated values are in good agreement, within the errors, among all studies. There is also general agreement between the different studies for He/H and He$^{+}$/H$^{+}$. However, He$^{2+}$/H$^{+}$ in the present study agrees well with PB08 and TS03, but is nearly twice as high compared to the estimates from MG13 and KC06, and about half the value reported by MH16. In their study, MH16 examined the chemical homogeneity of NGC\,3242 by splitting the main slit from the HST observations into nine smaller regions, determining the physicochemical properties of this PN in each specific region.
Using the specific slit analysis module of {\sc satellite}, ten pseudo-slits were replicated as illustrated in fig.~1 of MH16, to compare He/H and O/H in every region. For the total He abundance, our results agree well with MH16 within the errors. However, O/H values from MH16 are twice as high as our estimates.

\begin{table*}[h!]
    \centering
     \caption{
      Integrated $T_{\rm e}$,  $n_{\rm e}$, ionic and elemental abundances from this study in comparison with previous ones. }
     
    \begin{tabular}{|lcccccc|}

     \hline
      
     & Our study & MH16$^{\dag\dag\dag}$ & MG13 &  PB08 & KC06 &  TS03 \\
     
     \hline
     \multicolumn{7}{|c|}{$T_{\rm e}$(K)/$n_{\rm e}$ (cm$^{-3}$)}\\
     
     \hline
    $T_{\rm e}$ ({[S~\sc iii]}) & 12\,700$\pm$800 & & -& -& -&  - \\
     $T_{\rm e}$ ({[O~\sc iii]}) & -  & 11\,700$\pm$300 & 12\,900$\pm$720 & 10\,800 & 12\,140$\pm$31&  11\,700 \\
     $T_{\rm e}$ ({[N~\sc ii]}) & 11\,500$\pm$600 & - & 11\,670$\pm$1\,000 & 11\,000 & - & 13\,400 \\
     $n_{\rm e}$ ({[Cl~\sc iii]}) & 1\,600$\pm$900 & - &  2\,380$\pm$1\,300&  3\,000&  2\,531$\pm$1\,227& 1\,200 \\ 
     $n_{\rm e}$ ({[S~\sc ii]}) & 2\,300$\pm$700 & - &3\,120$\pm$1\,400&  1\,900 &1\,016$\pm$436 & 1\,970\\
     $n_{\rm e}$ ({C~\sc ii]}) & - & 4\,500$\pm$300 &-& - &- &-\\
    \hline

     \multicolumn{7}{|c|}{Ionic and elemental abundances}\\
     \hline

    He$^{+}$/H$^{+}$($\times$10$^{-2}$)& 7.35$\pm$0.42& 6.25$\pm$0.45& 8.0  & 7.07& 7.12$\pm$0.13 & 7.89\\
    He$^{2+}$/H$^{+}$(x10$^{-2}$)& 2.24$\pm$0.11   & 4.40$\pm$0.07  &0.945 & 2.12& 3.47$\pm$0.01 & 2.08\\
    He/H($\times$10$^{-2}$) &  9.60$\pm$0.47    &  10$\pm$5  &9  &  9.2 & 10.0$\pm$1.3 & 10\\
    \hline
    O$^{o}$/H$^{+}$($\times$10$^{-7}$) & 1.22$\pm$0.30 & - & 0.211&- & -&-\\
    O$^{+}$/H$^{+}$($\times$10$^{-6}$) & (6.7$\pm$1.6)$^\dag$ & (2.23$\pm$0.90)$^{\dag\dag}$& – & (5.1)$^{\dag\dag}$ & (1.4$\pm$0.1)$^{\dag\dag}$& 3.33\\
    O$^{2+}$/H$^{+}$($\times$10$^{-4}$)& 2.11$\pm$0.41  & 2.59$\pm$0.21 &2.34 & 3.78& 2.36$\pm$0.06 &  2.8\\
    O$^{3+}$/H$^{+}$($\times$10$^{-5}$)& -   & -&- & 4.3& -& 35.1 \\
    O/H($\times$10$^{-4}$)& (2.53$\pm$0.48)$^\dag$ & 4.44$\pm$0.36 &2.55 & 3.80 & 3.09$\pm$0.09& 3.3\\
    \hline
   ICF(O) & 1.2$\pm$0.3 & 1.70$\pm$0.05 &1.07 & 1 & 1.30$\pm$0.01 & 1.17\\
    \hline
    S$^{+}$/H$^{+}$($\times$10$^{-8}$) & 1.55$\pm$0.25   & -&0.602 & 1.80& 0.93$\pm$0.02 &  0.337\\
    S$^{2+}$/H$^{+}$ ($\times$10$^{-7}$) &  6.00$\pm$0.95  & - & 7.43 & 8.3& 6.1$\pm$0.2&  6.62\\
    S$^{3+}$/H$^{+}$($\times$10$^{-6}$)&   - &- &- & 1.94& -& -\\
    S/H($\times$10$^{-6}$)& (1.40$\pm$0.25)$^\dag$  & - & 2.24  & 2.80 & 2.60$\pm$0.11&  2.4\\
    \hline
   ICF(S) & (2.3$\pm$0.2)$^\dag$& - & 3.00   &1 & 4.20$\pm$0.11 & 3.52\\
    \hline
     N$^{o}$/H$^{+}$($\times$10$^{-8}$)& 2.32$\pm$0.95  &-  &-  &- & -&-\\
    N$^{+}$/H$^{+}$($\times$10$^{-7}$)&  (3.00$\pm$0.60)$^\dag$  & -&2.63& 4.1&1.9$\pm$0.1 & 3.2\\
     N$^{2+}$/H$^{+}$($\times$10$^{-5}$)&  -  & -&- & 5.7& - & 2.2 \\
     N$^{3+}$/H$^{+}$($\times$10$^{-6}$)& -   &- &- & 43.9&- & 6.77\\
      N$^{4+}$/H$^{+}$($\times$10$^{-5}$)&-    & -& -& 3.6&- & 2.35 \\
    N/H($\times$10$^{-5}$) & (1.50$\pm$0.41)$^\dag$   &- &2.05  & 13.5   & 4.19$\pm$0.39& 3.4 \\
    \hline
   ICF(N) & 49$\pm$14 &  -&80 & 1 & 220$\pm$17.0 & 1.53\\
    \hline
    \end{tabular}

    \tablefoot{The pseudo-slit employed for our estimations is centered at the center of the PN with dimensions 36.8$^{\prime\prime}$$\times$ 42.6 $^{\prime\prime}$. 
     \tablefoottext{b}{For the ionic abundances from previous studies, in the case where many emission lines were employed, and the adopted value was not available, the mean value of the individual ionic abundances was accounted in the table.}
     \tablefoottext {$\dag$} {The resulted values after the correction for recombination contribution. }
    \tablefoottext {${\dag\dag}$} {Only {[O~\sc ii]} $\lambda\lambda$3726,3729 (blended) were  available for the estimation of O$^{+}$/H$^{+}$.}
   \tablefoottext {${\dag\dag\dag}$}{ In the case of MH16, the values in the table are those that came from the pseudo-slit that covers the full region.}
  }
    \tablebib{
\citet{Miller_2016}:MH16, \citet{Monteiro2013}:MG13, \citet{Pottasch}:PB08, \citet{Krabbe}:KC06, \citet{Tsamis}:TS03.}\\
    
    \label{table_abund}
    
\end{table*}

Regarding sulphur abundance, we estimated  that it is (1.40$\pm$0.40)$\times$10$^{-6}$ while all previous studies found S/H from 2.2$\times$10$^{-6}$ to 2.8$\times$10$^{-6}$. A similar behavior was found for N/H, with estimates from the literature ranging from 2.05$\times$10$^{-5}$ to 13.5$\times$10$^{-5}$, which are between 37\% and 89\% higher than our estimated value. We further discuss the significance of these discrepancies in  Sect. \ref{sect_chemical_comp}.

\subsection{Diagnostic diagrams}
\label{subsec:DD}

Diagnostic diagrams (hereafter DDs) are designed to compare the relative line intensity ratios of certain emission lines commonly observed in PNe, Supernova Remnants (SNRs), and {H~\sc ii} regions. Each of these types of objects is expected to occupy a distinct region on the DDs, mainly due to their different excitation mechanisms. These diagrams are widely used to identify the aforementioned nebulae \citep[see,][]{frew_parker_2010, Sabin_diagnostics, Kopsacheili2020}. One of the most well-known DDs uses the ratio H$\rm\alpha$/[N {\sc ii}] versus  H$\rm\alpha$/{[S~\sc ii]} \citep{Sabbadin_diagnostics, diagnostics, leonidaki}. Another widely used DD relies on the {[O~\sc iii]}/H$\rm\beta$ versus {[S~\sc ii]}/H$\rm\alpha$ ratio, introduced by \cite{Baldwin_diagnostics}, to distinguish {H~\sc ii} regions, Seyfert galaxies, LINERS and PNe. DDs were initially developed from long-slit spectroscopic data. However, spatially resolved DDs have been used for the identification of shock excited regions or LISs in PNe \citep{Barria, Barria_ngc3242}, and DDs in conjunction with datacubes have also been used to identify SNRs \citep{ Monreal_Ibero2023, Kopsacheili2024}. Additionally, DDs have been employed to investigate the ionization structure and excitation mechanisms of PNe \citep{diagnostics_Akras,Belen_III}. 

To perform a direct comparison between 1D and 2D analysis, we ran the following {\sc satellite} modules: 2D analysis, specific slit analysis and rotational analysis, indicated as cyan dots, diamonds (colored based on the pseudo-slit's position) and pink dots on the DDs shown in Fig.~\ref{diagnostics}, respectively. For the specific slit analysis, four pseudo-slits were placed at the positions of the four knots (see, Fig.~\ref{emission_maps}), and another one covering the entire nebula was also used (pseudo-slit dimensions 36.8$^{\prime\prime}$$\times$42.6$^{\prime\prime}$). The rotational analysis was performed from P.A. 0\degr~to 360\degr, with 10\degr~steps. Since DDs were initially developed for 1D spectroscopic data, the results of the 2D analysis can only provide a glimpse into the distribution of each spaxel's value on these diagrams. This approach was previously used by \cite{diagnostics_Akras}, who analyzed the emission line ratios of each individual spaxel of Abell 14 on the DDs.

Overall, the cyan dots and the results from the slit analysis follow the same trend on these diagrams, indicating that the spaxel-by-spaxel representation on the DDs is well-defined. In the case of H$\rm\alpha$/{[N~\sc ii]} versus H$\rm\alpha$/{[S~\sc ii]} diagram (Fig.~\ref{diagnostics}, top panel) most of the cyan points lie outside the ellipse where PNe are expected to be, occupying a wide range of values from 0.5 to 3 for log(H$\rm\alpha$/{[N~\sc ii]} (6548+6584)) and from 1 to 3.5 for log(H$\rm\alpha$/{[S~\sc ii]} (6716+6731)). The brown ellipse marked on Fig.~\ref{diagnostics} is a density ellipse with 85\% probability (2 $\sigma$), which was derived from a sample of 613 PNe (light blue squares in Fig.~\ref{diagnostics}) \citep{diagnostics}.
To extend the region of PNe on this DD and account for high-ionization nebulae, such as NGC\,3242, we formulated covariance ellipse with 99\% probability (3 $\sigma$) based on the data points of NGC\,3242 (red dotted ellipse in Fig.~\ref{diagnostics}). It is worth noting that many spaxels from NGC~7009 have also been found outside the brown ellipse (see fig.~11 in \citealt{SATELLITE_case_studies}), as well as spaxels from 1D photoionization models with high log(U) parameter i.e. highly ionized nebulae (see fig.~5 in \citealt{diagnostics_Akras}). However, for a more general representation, a larger sample of high-ionization nebulae is needed and should be considered.
\footnote{log(U) is the logarithmic ratio of the ionizing photon density to hydrogen
density.}

A similar behavior is observed in the log([O \sc iii}] (4959+5007)/H$\rm\beta$) versus log({[N~\sc ii]} (6548+6584)/H$\rm\alpha$) DD. Although all points are within the PNe region, the values for the log({[N~\sc ii]} (6548+6584)/H$\rm\alpha$) ratio indicate that the intensity of H$\rm\alpha$ is significantly higher than that of [{N~\sc ii}]. These results reinforce the picture of a highly ionized PN \citep{excitation_ngc3242}, leading to higher values of the emission line ratios involved in the DDs. However, it should be noted that the values of the {[N~\sc ii]} (6548+6584)/H$\rm\alpha$ ratio are also affected by the metallicity of the nebula. The results from the specific slit analysis (colored diamonds) follow the same trend as the 2D and rotational representations. Only the pseudo-slits that are placed at the LISs (k1 and k4), exhibit higher log({[N~\sc ii]} (6548+6584)/H$\rm\alpha$), since the ionization degree there is, by definition, lower. 

Generally, caution is needed when these diagrams are used with spaxel values from IFU data due to possible degeneracies that may lead to misinterpretation of the results \citep[see a paradigmatic example presented by][]{Morisset2018}. The emission line ratios for the pseudo-slit that covers the whole PN (light-blue diamond), serve as indicators for the scattering of the spaxel-by-spaxel values. Specifically, log({[O~\sc iii]}$\lambda$5007/H$\rm\beta$) is 1.14, with spaxel values ranging from $-$0.04 to $+$0.39 ($\pm$ 0.02 for high flux-to-error (F/E) spaxels and $\pm$ 0.04 for low F/E spaxels)~\footnote{High F/E spaxels are considered those with H$\beta$ flux-to-error ratio above 70 and the rest of them are the low F/E spaxels. The first ones lye in the inner nebular regions and the others at the outer shell.}. The log(H$\rm\alpha$/{[S~\sc ii]}(6716+6731)) ratio has a central value of 2.64, with spaxel values spanning from $-$1.16 to $+$0.76 ($\pm$ 0.09 for high F/E and $\pm$ 0.10 for low F/E spaxels). For log(H$\rm\alpha$/{[N~\sc ii]}(6584+6548)), the ratio is 2.01, with deviations from $-$2.20 to $+$0.69 ($\pm$ 0.04 for high F/E and $\pm$ 0.05 for low F/E). Lastly, log({[N~\sc ii]}(6584)/H$\rm\alpha$) yields $-$2.14, with deviations ranging from $-$0.72 to $+$1.54 ($\pm$ 0.01 for high F/E and $\pm$ 0.02 for low F/E spaxels).

\begin{figure}
\centering
\includegraphics[scale=0.266]{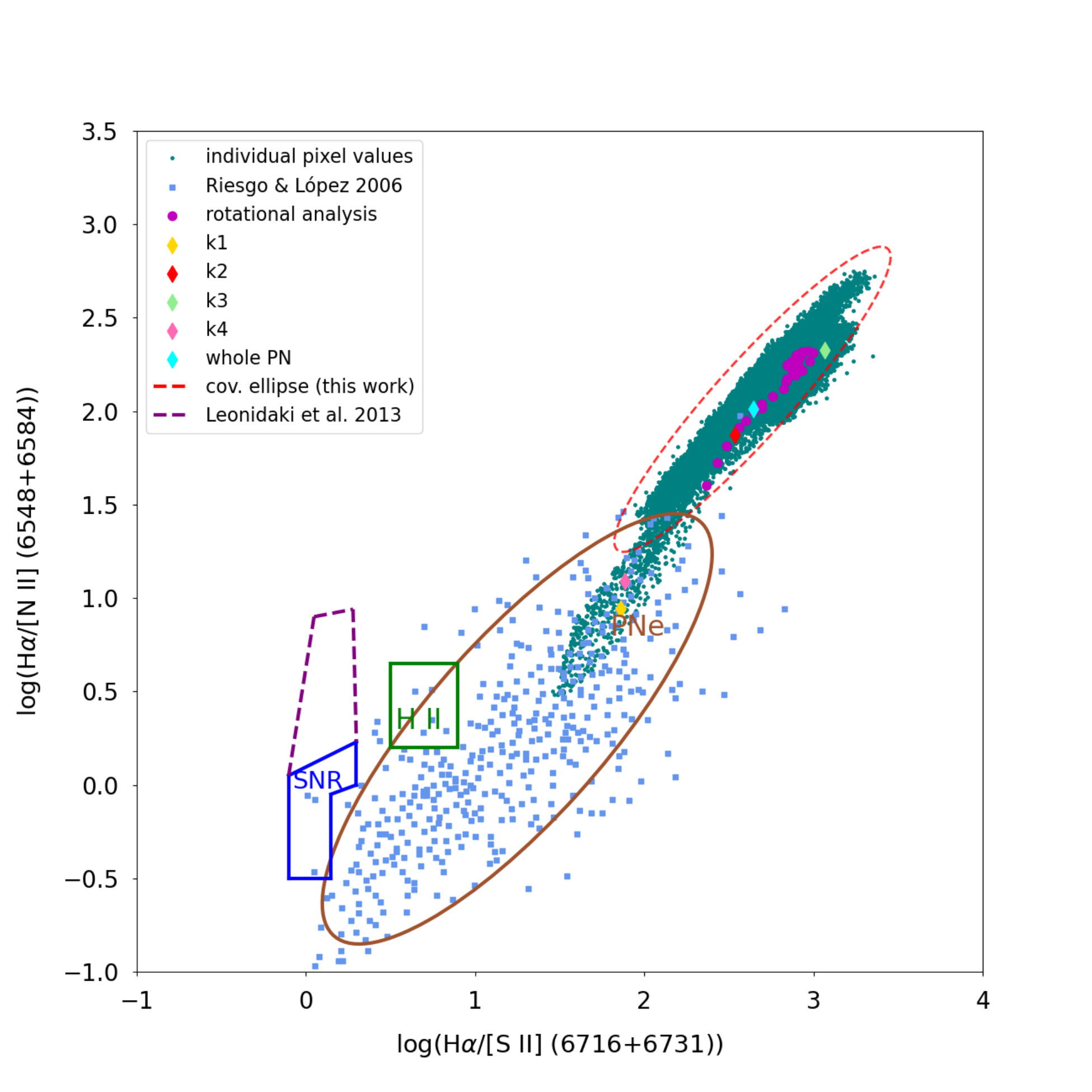}
\includegraphics[scale=0.266]{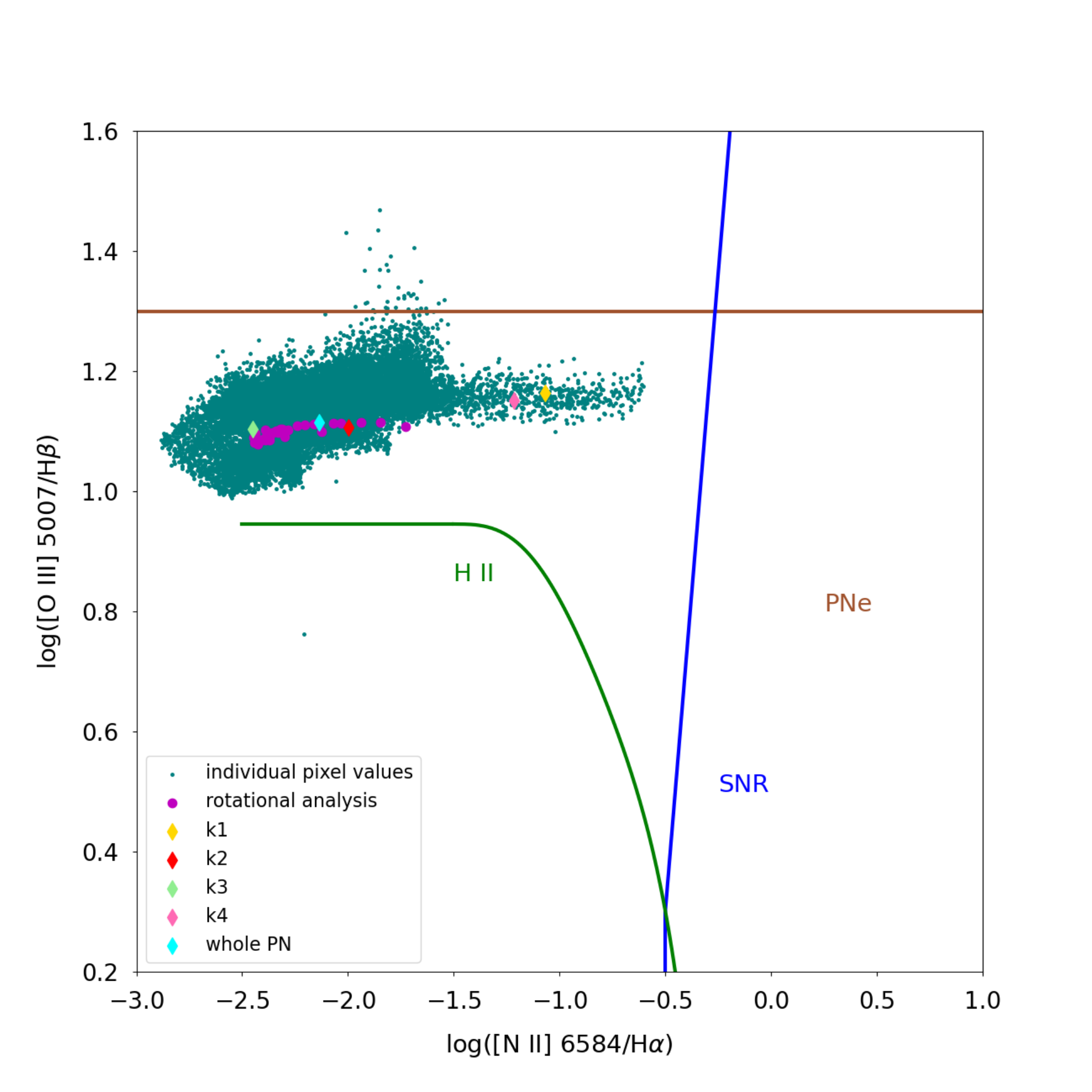}
\caption[]{Upper panel: log(H$\rm\alpha$/[N {\sc ii}] (6548+6584)) versus log(H$\rm\alpha$/{[S~\sc ii]} (6716+6731)). Lower panel: log([O {\sc iii}] (4959+5007)/H$\rm\beta$) versus log({[N~\sc ii]} (6548+6584)/H$\rm\alpha$). The brown, blue+purple and green regions define the positions of PNe, SNRs and {H~\sc ii} regions, respectively \citep{leonidaki, diagnostics}. The cyan dots are data points from the spaxel-by-spaxel analysis, the diamonds correspond to results from the specific slit analysis, and the pink dots indicate the findings from the rotational analysis. In the upper panel, the red dotted ellipse extends the PNe region to include the higher ionization PNe such as NGC\,3242.}
\label{diagnostics}
\end{figure}

\section{Discussion}
\label{sec:discussion}

\subsection{Nebular shell}

MUSE offers the capability to examine in detail the spatial distribution of various emission lines. The deeper investigation of MUSE datacubes led to the detection of new structures in the nebular shell. These regions are mainly seen in the light of {[S~\sc iii]} and {[N~\sc ii]}.
According to \cite{Ramos_Phillips_2009}, the fragmented halo of NGC\,3242, as revealed from Spitzer, is interacting with the ISM which could possibly cause the diffusion of ISM material into the nebular shells. Additionally, hydrodynamical instabilities in the inner nebular structures could yield to enhanced emission at the leading edges of the nebular shell. On the contrary, the {[Fe~\sc iii]} emission that has been detected in these structures, may indicate shock interactions (Bouvis et al., in prep.). However, the ratios of {[S~\sc ii]}/H$\alpha$ and {[N~\sc ii]}/H$\alpha$ reveal no trend for shock activity at these regions. So, further investigation is needed to understand their origin.

Two arc-like structures were also detected in the shell at the edges of the two LISs (see Fig.~\ref{emission_maps}). These could be related to the new structures found perpendicular to the LISs, since they share the same physico-chemical properties. However, the position of the arcs could also indicate an association with the LISs, but for now we have no evidence for this assumption.

\subsection{Physical parameters and excitation mechanisms}
\label{sec:discuss_phys}

In the present study, $T_{\rm e}$ was calculated from both CELs and ORLs. In every case, there is a trend toward higher temperatures in the inner nebular structures. $T_{\rm e}$ ([{S~\sc iii}]) and $T_{\rm e}$ ([{N~\sc ii}]) are systematically higher than $T_{\rm e}$ (PJ) and $T_{\rm e}$ ({He~\sc i}), with the latter being the lowest of all the estimated $T_{\rm e}$ values. The mean values for each $T_{\rm e}$ from the 2D temperature maps, with 1$\sigma$ confidence intervals, are $T_{\rm e}$ ([{S~\sc iii}]) = (12\,500$\pm$1\,700) K, $T_{\rm e}$ ([{N~\sc ii}]) = (11\,700$\pm$1\,500) K, $T_{\rm e}$ (PJ) = (8\,900$\pm$3\,700) K, and $T_{\rm e}$ ({He~\sc i}) = (7\,800$\pm$1\,000) K (see Table \ref{table_Te_Ne}). The fact that $T_{\rm e}$ is systematically higher when it is derived from CELs instead of ORL diagnostics, such as H~{\sc i} Balmer and/or Paschen discontinuities, is a well known issue. This discrepancy has been attributed whether to the presence of temperature inhomogeneities in the gas \citep{Peimbert}, or to the effect of chemically inhomogeneities \citep{Liu2000, Zhang2005nonlinear}. On the other hand, the behavior between $T_{\rm e}$ ({He~\sc i}) and $T_{\rm e}$ (PJ) is consistent with what was found for a sample of 48 PNe by \citet{Zhang2005nonlinear} ($T_{\rm e}$ ({He~\sc i}) $<$ $T_{\rm e}$ (H~{\sc i})). These authors considered this behavior to be consistent with the expectations of a chemically inhomogeneous nebula model. However, in a recent paper, \citet{MendezDelgado2024} have found that a combination of temperature inhomogeneities and deviations from ``case-B'' recombination scenario can explain the observed discrepancies between $T_{\rm e}$ ({He~\sc i}) and other temperature diagnostics. It is not trivial to assess the contribution of each scenario to this discrepancy. \citet{MendezDelgado2024} propose a toy model with a mixed ``case A'' + ``case B'' scenario to check if deviations from ``case B'' could account for the temperature discrepancies, by using the fact that ``case A'' and ``case B'' He~{\sc i} line emissivities originating from $n^1P$ levels differ substantially. Unfortunately, the only transition originating from a $n^1P$ level in the wavelength range covered by MUSE is He~{\sc i} $\lambda$5016, whose emission line map in the long-exposure datacube could not be extracted owing to the proximity of [O~{\sc iii}] $\lambda$5007 line, which is strongly saturated; alternatively, we tried to extract it in the 10s exposure datacube, but the obtained map was too noisy to draw any significant conclusion.

Regarding the electron density, a trend toward higher values is observed at the nebular rim, while no significant difference is observed between $n_{\rm e}$ ([{S~\sc ii}]) = (2\,200$\pm$1\,500) cm$^{-3}$ and $n_{\rm e}$ ([{Cl~\sc iii}]) = (1\,600$\pm$1\,000) cm$^{-3}$ (see Table \ref{table_Te_Ne}).

\subsection{Nebular chemical composition}
\label{sect_chemical_comp}

The chemical properties of NGC\,3242 estimated from this study are generally in good agreement with previous ones. However,  there is a trend to higher values for O/H with the biggest difference arising from MH16. The value of MH16 for the entire region (included in Table~\ref{table_abund}) is closer to the value reported by PB08, but still higher. It is worth mentioning though that 
MH16 used the recipe of \cite{Kwitter2001} for the ICF(O). Based on this approach, ICF is a linear representation of the (He$^{+}$+ He$^{2+}$)/ He$^{+}$ ratio, in contrast to the ICF(O) that is employed in this work, which depends on a power of 2/3 of the same ratio \citep[see eq. A9 in][]{KB}.

Also, the discrepancy in S abundance among our study with the previous ones, is significant. We suggest that the difference arises from the lower ionization degree (O$^+$/O) obtained from our data compared to the literature values, which translates into 
a lower ICF. As mentioned above, PB08 took advantage of a multi-wavelength analysis and thus do not adopt any ICF to compute the total S abundance, which is $\sim$50\% higher than the one computed in this work.

Similarly, N/H is notably higher, than our estimation (Table \ref{table_abund}). However, the computation of total N abundances from optical spectra in high-excitation PNe is controversial, because of the extremely large and uncertain corrections that should be adopted. PB08 computed the N abundance without adopting an ICF, by combining N$^+$ ionic abundances from the optical and N$^{2+}$, N$^{3+}$, and N$^{4+}$ from ultraviolet (UV) and far-infrared observations (see values in Table \ref{table_abund}). With these data they estimated a log (N/O) $\sim -0.44$, which is much higher than the N/O ratios estimated from other literature sources (including TS03 which also used UV and far-infrared observations to compute N$^{2+}$, and N$^{3+}$), that imply log (N/O) ratios between $-0.86$ and $-1.14$). However, it is worth mentioning that TS03 and PB08 obtained very different N$^{2+}$, and N$^{3+}$ abundances from different sets of UV and far-infrared data and that several of these values were reported by these authors as highly uncertain. Moreover, N$^{2+}$ abundance from the far-infrared N~{\sc iii}] 57.3$\mu$m line is very density-dependent, finding differences up to a factor of 2, depending on the adopted $n_{\rm e}$ (see discussion in sect.~5.2 from TS03).

\subsection{The Low Ionization Structures}

\subsubsection{Emission of atomic carbon from the LISs}
LISs are small scale structures with strong emission in low ionization species, compared to the surrounding nebula. Commonly, {[N~\sc ii]}, {[S~\sc ii]} and {[O~\sc i]} are identified in the LISs of PNe. In this study, the investigation of MUSE data led to the detection of the {[C~\sc i]} $\lambda$8727 emission line, too. The same line of neutral C has also been detected in the LISs of NGC\,7009 \citep{Akras_2024}. In addition, atomic O is also  mainly emitted from the pair of LISs, (see Fig.~\ref{CI_OI} right panels). This co-existence has been previously noticed in the knotty structures of NGC\,6778, and M\,1-42 \citep{ADF_recomb}. The co-spatial distribution of atomic C and atomic O emissions, may imply the dissociation of CO due to the intense UV radiation from the CS, but further investigation is needed (Gonçalves et al. in preparation). CO has been previously identified in the cometary knots of Helix nebula \citep{CO_Helix, Helix2020}, while \cite{Baez2023}, using ALMA observations, detected CO in the clumpy structures of several PNe.

Molecular hydrogen (H$_2$) emission has also been found to originate in LISs \citep{AKras_2017, LIS_H2_Akras, Fang_2015, Fang_2018}. In the case of NGC~7009, the ionization stratification was found to be equivalent to a mini-PDR surrounding a molecular/H$_2$ core \citep{LIS_H2_Akras, Akras_2024}. More specific, the high ionization lines peak closer to the CS than the low ionization ones, and then the atomic lines follow the sequence. Even further, in the case of NGC~7009, molecular lines of H$_2$ were found. In the present study on NGC\,3242, the radial analysis indeed revealed that the high ionization lines (H$\rm\alpha$, H$\rm\beta$, {[O~\sc iii]}, {He~\sc ii} and {[S~\sc iii]}) peak at the nebular rim while the low ionization ({[N~\sc ii]} and {[S~\sc ii]}) and neutral ({[O~\sc i]} and {[C~\sc i]}) lines peak at the LISs. Generally, there is no significant offset between the peaks of the low ionization lines. Interestingly, though, in the case of the high-to-moderate ionization lines, a different stratification is observed among the two P.A. We suggest, that this divergence more likely occurs due to the orientation of the nebula and the shock emission due to the interaction of the inner jet with the rim (see Section \ref{Iron_emission}), perplexing, the surface brightnesses profiles.

\subsubsection{Search for H$_2$ in Spitzer Data}
In the LISs of NGC\,3242, $n_{\rm e}$ was found lower compared to the main nebular structures, while theoretical models predict the exact opposite. This problem could be ruled out if molecular gas exists in LISs of NGC\,3242, such as the low ionization features of other PNe \citep{Speck_2003, Kwok_2008, Matsuura_2009_Helix, Fang_2015, AKras_2017, Fang_2018, LIS_H2_Akras, Wesson_2023}, but until now there is no direct evidence for this.
In an attempt to further investigate the assumption of H$_2$ presence at the LISs, archival data from Spitzer Space Telescope (SST) for NGC\,3242 were utilized, in order to search for H$_2$ in Infrared Array Camera (IRAC) bands. IRAC is a four-channel camera that provided simultaneous 5.2$^\prime$ $\times$ 5.2$^\prime$ images at 3.6, 4.5, 5.8, and 8.0 $\mu$m. The filters had bandwidths $\Delta\lambda = 0.75~\mu$m, $\Delta\lambda = 1.902~\mu$m, $\Delta\lambda = 1.425~\mu$m and $\Delta\lambda = 2.905~\mu$m.   At the position of the LISs in the planetary nebula NGC\,7009, the ratios of the IRAC filters with the 4.5 $\mu$m were decreasing \citep{Phillips_H2}. It was suggested that this decrease is associated with an increase in emission from the 4.5 $\mu$m band. \cite{LIS_H2_Akras} proposed that the primary emission in this band is likely H$_2$. However, in the case of NGC\,3242, the LISs were not detectable in any of IRAC bands. This lack of observation suggests that the emission from H$_2$ is either faint, or absent. 

A correlation between the intensities of {[O~\sc i]} $\lambda$6300 line and the l-0 S(1) ro-vibrational line of H$_2$, was introduced by \cite{Reay}. The results about a sample of ten PNe is presented in their figure 2. Based on that plot, we can have an indication about the flux of H$_2$ given the flux of {[O~\sc i]}. In the case of NGC\,3242, the flux of {[O~\sc i]} $\lambda$6300 line integrated at the LISs is on the order of 10$^{-14}$ erg\, cm$^{-2}$\,s$^{-1}$, which implies a flux of H$_2$ around 10$^{-16}$  erg\, cm$^{-2}$\,s$^{-1}$. This flux is translated in a flux density of  10$^{-32}$  erg\, cm$^{-2}$\,s$^{-1}$\,Hz$^{-1}$, given the bandwidth of the filter and the exposure time of the observations. In the same exposure time, the sensitivity of IRAC in the 4.5 $\mu$m band, reaches a sensitivity of $\sim$10$^{-29}$  erg\, cm$^{-2}$\,s$^{-1}$\,Hz$^{-1}$. So, if indeed there is emission of H$_2$, it would be really difficult to be detected from broadband filters like the ones provided from IRAC.

\subsubsection{{[Fe\sc~ii]} and {[Fe\sc~iii]} emission from the LISs}

In addition to the detection of the {[Fe\sc~ii]} $\lambda$8617 line from the LISs, the {[Fe\sc~iii]} $\lambda$5270 line was also detected, showing the presence of a jet-like structure that connects the central star and the LISs. In general, iron is considered as shock indicator, so in the case of NGC~3242 Fe is probably released and ionized from shocks too. However, we further investigate the origin of iron emission in a follow-up study (Bouvis et al. in prep.). Moreover, the formation of a jet could indicate the existence of a binary progenitor, consisting of a companion star rotating around an AGB star \citep{Soker1994}. The scenario of a multiple star system, at the center of NGC~3242, has been previously reported from \cite{Soker1992}.  It is interesting though that the jet is asymmetric, with the northwest part being more extended. The X-ray emission, found in NGC\,3242, also appears stronger in the same direction \citep{Ruiz2011}. However, from Fig.~\ref{velocity}, it is clear that k1 is red-shifted while k4 is blue-shifted, so it is possible that the orientation of the PN is responsible for both the asymmetry of the jet and the enhanced X-ray emission at the northwest nebular regions.

The interaction of the jet with the rim may be responsible for the formation of the blobs (b1 and b2) and possibly the knots. In particular, k2 and k3 structures are characterized by higher {[O\sc~iii]}/{[O\sc~i]} and {[S\sc~iii]}/{[S\sc~ii]} ratios compared to the host nebula. So, the enhanced emission of {[O\sc~iii]} and {[S\sc~iii]} at the rim, and the formation of knotty structures along the direction of the LISs, could also be an aftermath of the jet-rim interaction. Nevertheless, there is no discrepancy in $T_{\rm e}$ among the different regions (Table \ref{table_shell_knots}). Concluding, we argue that there is a correlation between the knots and the newly discovered jet-like structure, which may also imply their shock excitation. Although this seems to contradict the lower $n_{\rm e}$ found for k1 and k4,  the higher densities expected in shock environments could be explained by a molecular component, as $n_{\rm e}$ reflects only the ionized gas fraction. We further examine the dominant excitation mechanism of k1 and k4 in an upcoming paper (Bouvis et al. in prep.).

\section{Conclusions}
\label{sec:summary}

In this study, we investigated the physico-chemical properties of NGC\,3242 using MUSE IFU data in conjunction with the {\sc satellite} code. A detailed examination of the emission line maps, allowed us to identify new structures embedded in the nebular shell, oriented perpendicular to the pair of LISs. These structures may represent the aftermath of instabilities in the inner nebular structures. However, we cannot rule out the possibility of shock interactions. Two arc-like structures were also identified at the edges of the LISs. We suggest that these are probably linked to the enhanced emission of the nebular shell, but they could also be associated with the pair of LISs. The extinction coefficient (c(H$\beta$)), calculated from H~{\sc i} Balmer lines, shows no significant variation throughout the nebula, with a mean value of 0.14. $T_{\rm e}$({[N~\sc ii]}) and $T_{\rm e}$({[S~\sc iii]}) are  approximately  11\,8700 K and 12\,500 K, respectively, while $n_{\rm e}$ ({[S~\sc ii]}) and $n_{\rm e}$ ({[Cl~\sc iii]}) are around 2\,200 and 1\,600 cm$^{-3}$, respectively.
Regarding  $T_{\rm e}$ from ORLs, $T_{\rm e}$ (PJ) is 8\,900 and $T_{\rm e}$(He {\sc~i}) is 7\,800 K, in an ellipse that covers the inner nebular structures to avoid the contamination from the  $T_{\rm e}$ outside the rim. Unfortunately, the data were not deep enough to test whether the discrepancy between $T_{\rm e}$ (PJ) and $T_{\rm e}$ (He {\sc i}) could be the aftermath of deviations from the ``case B''.

The diagnostic diagrams provided by {\sc satellite}, revealed that NGC\,3242 is a highly ionized nebula, and its dominant excitation mechanism is photoionization. In terms of the chemical composition of this PN, there is no significant variation throughout the nebular structures. Our results are consistent with previous studies, except for the elemental abundances of O, S and N. For O/H, the biggest difference arises in the case of MH16 where different ICF(O) recipe has been used.
Regarding, N/H and S/H, since NGC\,3242 is a highly ionized PN, the contribution from triply ionized N and S lines, which are not available in optical data, is crucial.

The {[C~\sc i]} $\lambda$8727 atomic line was identified in MUSE data, originating mainly from the LISs. The co-existence of {[C~\sc i]} and {[O~\sc i]} at the LISs may suggest the dissociation of CO molecules, although, until now, there is no direct confirmation of this hypothesis. Additionally, through the radial analysis module, we found that the atomic lines of {[C~\sc i]} and {[O~\sc i]} as well as other low-ionization lines peak closer to the LISs compared to moderate to high-ionization lines. Thus, LISs may consist of a molecular core surrounded by highly ionized gas, which is further enclosed by a partially ionized gas. In the present study, we were unable to detect H$_2$ through Spitzer IRAC data, but further investigation is needed to confirm the existence or absence of H$_2$ at the LISs of NGC\,3242. 

Moreover, iron emission was detected in the spectrum of NGC\,3242. More precisely, {[Fe~\sc ii]} $\lambda$8617 was found to emanate from the LISs and {[Fe~\sc iii]} $\lambda$5270 was detected in a jet-like structure that extends from the center of the nebula to the LISs. This finding may suggest that the knots and the jet-like structure are correlated. We also argue that both {[Fe~\sc ii]} and {[Fe~\sc iii]} are probably shock originated, while the presence of the jet could also indicate the existence of a binary progenitor.

\begin{acknowledgements}
The research project is implemented in the framework of H.F.R.I call “Basic research financing (Horizontal support of all Sciences)” under the National Recovery and Resilience
Plan “Greece 2.0” funded by the European Union–NextGenerationEU (H.F.R.I. Project Number: 15665). JGR acknowledges financial support from the Agencia Estatal de Investigaci\'on of the Ministerio de Ciencia e Innovaci\'on (AEI- MCINN) under Severo Ochoa centers of Excellence Programme 2020-2023 (CEX2019-000920-S), and from grant ``Planetary nebulae as the key to understanding binary stellar evolution'' with reference number PID-2022-136653NA-I00 (DOI:10.13039/501100011033) funded by the Ministerio de Ciencia, Innovación y Universidades (MCIU/AEI) and by ERDF "A way of making Europe" of the European Union. The authors would like to acknowledge Dr. Henri Boffin for the reduction of the MUSE data. The following software packages in Python were used: Matplotlib \citep{Hunter2007}, NumPy \citep{Walt2011}, SciPy \citep{SciPy2020} and AstroPy Python \citep{Astropy2013,Astropy2018}.

\end{acknowledgements}
%
%

\bibliographystyle{aa} 
\bibliography{biblio} 
\begin{appendix}

\section{Supplementary tables}

\begin{table}[h!]
    \caption{Observed and de-reddened line fluxes relative to H$\beta$=100.}
    \centering
    \resizebox{0.3\textwidth}{!}{
        \begin{tabular}{|lcc|}
            \hline
            Line ($\AA$) & F($\lambda$) & I($\lambda$) \\
            \hline
            H$\beta$ & 100 & 100\\
            {[O\sc~iii]} 4959 & 435.0$\pm$8.7  &  432.0$\pm$8.7\\
            {[O\sc~iii]} 5007 &1\,314$\pm$26.4 & 1\,299.0$\pm$26.5\\
            {[N\sc~i]} 5200 & 0.016$\pm$0.002& 0.016$\pm$0.002\\
            He{\sc~ii} 5412& 2.14$\pm$0.11 &  2.05$\pm$0.1\\
            {[Cl\sc~iii]} 5517 &0.032$\pm$0.03  & 0.31$\pm$0.03\\
            {[Cl\sc~iii]} 5538 & 0.30$\pm$0.03 & 0.28$\pm$0.03\\
            {[N\sc~ii]} 5755$^{\dag}$ & 0.052$\pm$0.007 & 0.064$\pm$0.007\\
            & 0.032$\pm$0.03 &  0.049$\pm$0.005\\
            He{\sc~i} 5876& 11.97$\pm$0.63 & 12.0$\pm$0.6\\
            {[O\sc~i]} 6300 & 0.105$\pm$0.011 & 0.096$\pm$0.010\\
            {[S\sc~iii]} 6312 & 0.76$\pm$0.08 & 0.70$\pm$0.08\\ 
            {[O\sc~i]} 6363 & 0.041$\pm$0.001 & 0.04$\pm$0.01\\
            {[N\sc~ii]} 6548 & 0.76$\pm$0.04 &  0.70$\pm$0.04\\ 
            H$\alpha$ & 319.0$\pm$ 6.4&  287.0$\pm$6.4\\ 
            {[N\sc~ii]} 6584 & 2.33$\pm$0.12 & 2.10$\pm$0.12\\
            He{\sc~i} 6678 & 3.30$\pm$0.17 & 3.0$\pm$0.21\\ 
            {[S\sc~ii]} 6716 & 0.31$\pm$0.02&  0.27$\pm$0.02\\
            {[S\sc~ii]} 6731 &  0.42$\pm$0.02& 0.37$\pm$0.02\\
            {[Ar\sc~iii]} 7136 & 8.77$\pm$0.46&  7.7$\pm$0.5\\
            {[O\sc~ii]} 7320$^{\dag}$ & 0.63$\pm$0.07& 0.55$\pm$0.06\\ 
            & 0.54$\pm$0.06 &  0.46$\pm$0.07\\ 
            {[O\sc~ii]} 7330$^{\dag}$ & 0.56$\pm$0.06 &0.50$\pm$0.06\\ 
            & 0.45$\pm$0.05 &  0.38$\pm$0.07\\ 
            {[C\sc~i]} 8727 & 0.008$\pm$0.0001 & 0.0070$\pm$0.0001 \\
            {[S\sc~iii]} 9069 & 7.6$\pm$0.4&  6.3$\pm$0.5\\ 
            F(H$\beta$)$\times$10$^{-14}$ & 140.0$\pm$1.4 &  - \\ \hline
            c(H$\beta$)  & \multicolumn{2}{c|}{0.14}  \\ \hline
        \end{tabular}
    }

    \tablefoot{ The fluxes are integrated for a pseudo-slit that covers the whole PN (36.8$^{\prime\prime}$$\times$ 42.6 $^{\prime\prime}$ and centered at the center of the PN). 
    \tablefoottext{a} {For [N~{\sc ii}] $\lambda$5755, the flux was estimated from the raw map} 
    \tablefoottext{b} F(H$\beta$) is the reddened H$\beta$ flux in unit erg\,cm$^{-2}$\,s$^{-1}$. 
    \tablefoottext{$\dag$} {The second row represents the values after correction for recombination contribution.}
    }

    \label{line_fluxes}
\end{table}

\begin{table}[h!]
\centering
\caption{Physical parameters and emission line ratios, resulting from the specific slit analysis for the new structures (2$^{nd}$ column), the four knots (columns 3 to 6) and the surrounding environment of the new structures (final column).}
\label{table_shell_knots}

\resizebox{0.7\textwidth}{!}{  
    \begin{tabular}{|lcccccc|}  
        \hline
        & new structures & k1 & k2 & k3 & k4 & surrounding gas  \\
        \hline
        \multicolumn{7}{|c|}{$T_{\rm e}$(K)/$n_{\rm e}$ (cm$^{-3}$)}\\
        \hline
        $T_{\rm e}$ ({[S~\sc iii]}) & 11\,600$\pm$660&  11\,700$\pm$660& 12\,100$\pm$760 & 14\,900$\pm$1\,100  & 11\,700$\pm$700 & 11\,900$\pm$790 \\
        $T_{\rm e}$ ({[N~\sc ii]})&  12\,000$\pm$770& 11\,300$\pm$500& 10\,700$\pm$600 & 11\,400$\pm$700  & 11\,400$\pm$600 & 11\,700$\pm$840 \\
        $n_{\rm e}$ ({[Cl~\sc iii]}) & 900$\pm$500 & 1\,200$\pm$870& 3\,100$\pm$1\,300  & 2\,300$\pm$1\,100  & 1\,400$\pm$800& 900$\pm$300\\ 
        $n_{\rm e}$ ({[S~\sc ii]}) & 1\,500$\pm$300 & 1\,800$\pm$500& 4\,500$\pm$1\,000& 3\,400$\pm$1\,000& 2\,200$\pm$500 & 1\,500$\pm$300\\
        \hline
        \multicolumn{7}{|c|}{Emission line ratios}\\
        \hline
        log ({[\sc S iii]}/{[\sc S ii]}) & 1.06$\pm$0.04&  0.37$\pm$0.04 & 1.01$\pm$0.03 & 1.25$\pm$0.04 & 0.58$\pm$0.04&1.16$\pm$0.04\\
        log ({[\sc O iii]}/{[\sc O i]}) & 4.18$\pm$0.04&  2.60$\pm$0.04 & 4.80$\pm$0.03& 4.74$\pm$0.04&2.97$\pm$0.04&4.22$\pm$0.04\\
        log ({[\sc N ii]}/H$\rm\alpha$) & $-$2.04$\pm$0.03&$-$1.00$\pm$0.02&$-$1.90$\pm$0.02& $-$2.30$\pm$0.02& $-$1.20$\pm$0.02& $-$2.30$\pm$0.03\\
        log ({[\sc S ii]}/H$\rm\alpha$) & $-$2.54$\pm$0.03 & $-$1.86 $\pm$0.02 &$-$2.54$\pm$0.02& $-$3.70$\pm$0.03 &$-$2.10$\pm$0.02 & $-$2.80$\pm$0.03\\
        log ({[N~\sc ii]}/{[\sc O iii]}) &$-$2.85$\pm$0.02 &$-$1.88$\pm$0.02&$-$2.69$\pm$0.02& $-$3.05$\pm$0.02 &$-$1.85$\pm$0.02 & $-$3.10$\pm$0.02\\
        \hline
    \end{tabular}
}

\tablefoot{Where {[S~\sc iii]}, {[S~\sc ii]}, {[O~\sc iii]}, {[O~\sc i]} and {[N~\sc ii]} stand for {[S~\sc iii]} $\lambda$(6312+9069), {[S~\sc ii]} $\lambda$(6716+6731), {[O~\sc iii]} $\lambda$(4959+5007), {[O~\sc i]} $\lambda$(6300+6363) and {[N~\sc ii]} $\lambda$(6548+6583), respectively.}
\end{table}

\section{2D abundance maps}

\begin{figure}[h!]
    \centering
    \null\hfill
    \subfloat{\includegraphics[width =0.4\textwidth]{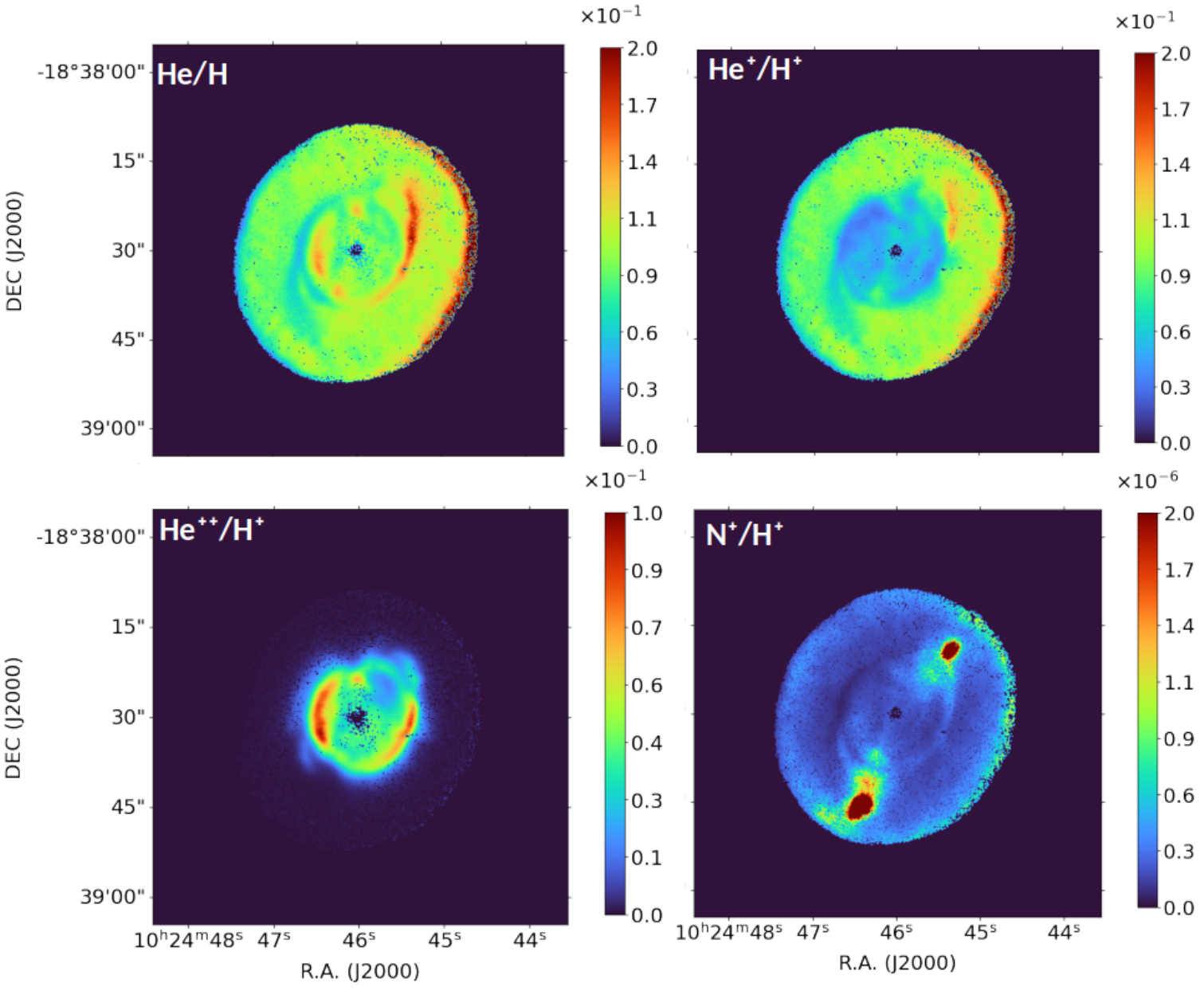}}
    \hfill
    \caption{Ionic and elemental abundance maps. From left to right and from top to bottom: He/H, He$^{+}$/H$^{+}$, He$^{2+}$/H$^{+}$, N$^{+}$/H$^{+}$.}
    \hfill
\label{abund_1}
\end{figure}

\begin{figure}[h!]
    \centering
    \null\hfill
    \subfloat{\includegraphics[width =0.4\textwidth]{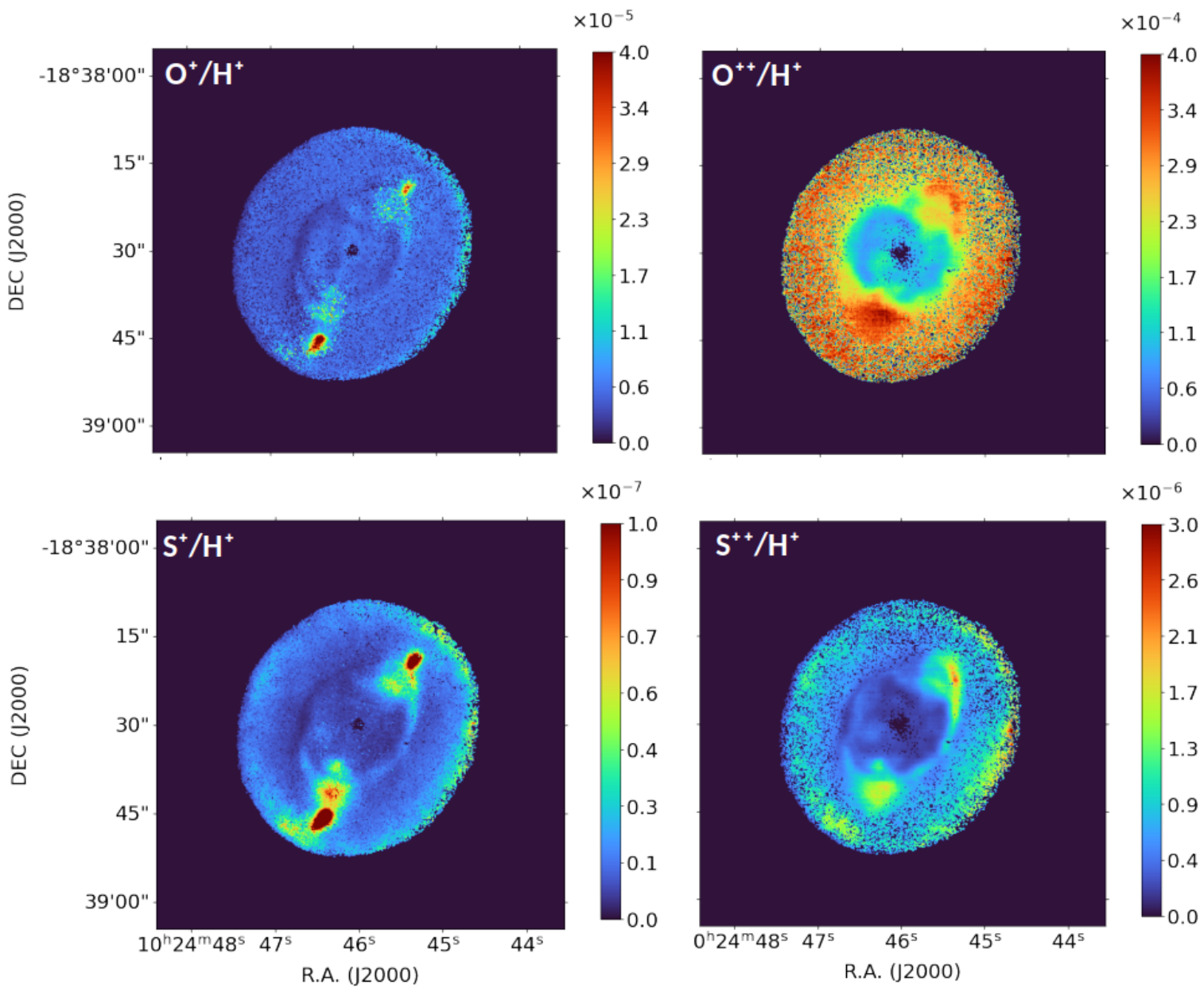}}
    \hfill
    \caption{Ionic and elemental abundance maps. From left to right and from top to bottom: O$^{+}$/H$^{+}$, O$^{2+}$/H$^{+}$, S$^{+}$/H$^{+}$, and S$^{2+}$/H$^{+}$.}
    \hfill
\label{abund_2}
\end{figure}

\end{appendix}

\end{document}